\documentclass[a4paper,twocolumn,showpacs,superscriptaddress,floatfix,nofootinbib,prd]{revtex4-1}
\usepackage{graphicx}
\usepackage{epsf}
\usepackage{epsfig}
\usepackage{amssymb,amsmath}
\usepackage[usenames]{color}
\usepackage{amssymb}
\usepackage{times}
\usepackage{mathrsfs}
\usepackage{hyperref}
\hypersetup{
  colorlinks=true,        % false: boxed links; true: colored links
  linkcolor=blue,         % color of internal links
  citecolor=cyan,         % color of links to bibliography
}
\usepackage{float}
\newcommand{\lrn}[1]{\textcolor{black}{#1}}

\definecolor{orange}{rgb}{1,0.5,0}

\newcommand{\cf}{cf.~}
\newcommand{\ie}{i.e.,~}
\newcommand{\eg}{e.g.,~}

\renewcommand{\BibitemShut}[1]{}
\newcommand{\ms}{\,{\rm ms}}
\newcommand{\km}{\,{\mathrm{km}}}
\newcommand{\gcm}{\,{\mathrm{g}/\mathrm{cm}^{3}}}
\newcommand{\Msun}{M_{\odot}}
\begin{document}

\title[Rotational properties of HMNSs from binary mergers]{Rotational
  properties of hypermassive neutron stars from binary mergers}

\author{Matthias~Hanauske}
\affiliation{Institut f{\"u}r Theoretische Physik, Max-von-Laue-Stra{\ss}e 1, 60438 Frankfurt, Germany}
\affiliation{Frankfurt Institute for Advanced Studies, Ruth-Moufang-Stra{\ss}e 1, 60438 Frankfurt, Germany}
%%%
\author{Kentaro~Takami}
\affiliation{Kobe City College of Technology, 651-2194 Kobe, Japan}
\affiliation{Institut f{\"u}r Theoretische Physik, Max-von-Laue-Stra{\ss}e 1, 60438 Frankfurt, Germany}
%% 
%%%
%% \author{Bruno~Mundim}
%% \affiliation{Institut f{\"u}r Theoretische Physik, Max-von-Laue-Stra{\ss}e 1, 60438 Frankfurt, Germany}
%%%
\author{Luke~Bovard}
\affiliation{Institut f{\"u}r Theoretische Physik, Max-von-Laue-Stra{\ss}e 1, 60438 Frankfurt, Germany}
%%%
\author{Luciano~Rezzolla}
\affiliation{Institut f{\"u}r Theoretische Physik, Max-von-Laue-Stra{\ss}e 1, 60438 Frankfurt, Germany}
\affiliation{Frankfurt Institute for Advanced Studies, Ruth-Moufang-Stra{\ss}e 1, 60438 Frankfurt, Germany}
%%%
\author{Jos\'{e}~A.~Font}
\affiliation{Departamento de Astronom\'{\i}a y Astrof\'{\i}sica, Universitat de Val\`encia, Dr. Moliner 50, 46100, Burjassot (Val\`encia), Spain}
\affiliation{Observatori Astron\`omic, Universitat de Val\`encia, C/ Catedr\'atico Jos\'e Beltr\'an 2, 46980, Paterna (Val\`encia), Spain}
%%%
\author{Filippo~Galeazzi}
\affiliation{Institut f{\"u}r Theoretische Physik, Max-von-Laue-Stra{\ss}e 1, 60438 Frankfurt, Germany}
%%%
\author{Horst~St\"ocker}
\affiliation{Institut f{\"u}r Theoretische Physik, Max-von-Laue-Stra{\ss}e 1, 60438 Frankfurt, Germany}
\affiliation{Frankfurt Institute for Advanced Studies, Ruth-Moufang-Stra{\ss}e 1, 60438 Frankfurt, Germany}
\affiliation{GSI Helmholtzzentrum f{\"u}r Schwerionenforschung GmbH, 64291 Darmstadt, Germany}
%%%

\begin{abstract}
Determining the differential-rotation law of compact stellar objects
produced in binary neutron stars mergers or core-collapse supernovae is
an old problem in relativistic astrophysics. Addressing this problem is
important because it impacts directly on the maximum mass these objects
can attain and hence on the threshold to black-hole formation under
realistic conditions.
Using the results from a large number of numerical simulations in full
general relativity of binary neutron star mergers described with various
equations of state and masses, we study the rotational properties of the
resulting hypermassive neutron stars.
We find that the angular-velocity distribution shows only a modest
dependence on the equation of state, thus exhibiting the traits of
``quasi-universality'' found in other aspects of compact stars, both
isolated and in binary systems.  The distributions are characterized by
an almost uniformly rotating core and a 
``disk''. Such a configuration is significantly different from the
$j-{\rm constant}$ differential-rotation law that is commonly adopted in
equilibrium models of differentially rotating stars.
Furthermore, the rest-mass contained in such a disk can be quite large,
ranging from $\simeq 0.03\,M_{\odot}$ in the case of high-mass binaries
with stiff equations of state, up to $\simeq 0.2\,M_{\odot}$ for low-mass
binaries with soft equations of state.  We comment on the
astrophysical implications of our findings and on the long-term
evolutionary scenarios that can be conjectured on the basis of our
simulations.
\end{abstract}


\maketitle

%%%%%%%%%%%%%%%%%%%%%%%%%%%%%%%%%%%%%%%%%%%%%%%%%%%%%%%%%%%%%%%%%%
\section{Introduction}
\label{sec:intro}
%%%%%%%%%%%%%%%%%%%%%%%%%%%%%%%%%%%%%%%%%%%%%%%%%%%%%%%%%%%%%%%%%%

A number of catastrophic astrophysical events, such as the collapse of
the iron core in Type-II supernovae or the merger of a binary system of
neutron stars, lead to the formation of a compact object with a large
amount of angular momentum. Clearly, uniform rotation is not an efficient
way of sustaining such large rotation rates and it is far easier to
obtain an equilibrium by distributing the angular momentum differentially
in radius. At the same time, for a given amount of mass involved in the
event, the knowledge of the law of differential rotation is important in
establishing how close or how far the equilibrium configuration attained
is from the instability threshold of the gravitational collapse to a
black hole. On the basis of these considerations, it is clear that the
problem of determining the law of differential rotation produced in these
events is an important one.

Over the last 15 years, a large number of works have explored this
problem in full general relativity, either through the construction of
equilibrium configurations \cite{Baumgarte00bb, Lyford2003, Shibata05c,
  Shibata06a, Ansorg2009, Galeazzi:2011, Studzinska2016, Gondek2016} or
through their dynamical production in core-collapse supernovae (see
\cite{Ott06c} for an overview) or in binary neutron star mergers (see
\cite{Baiotti2016} for a recent review). Lacking a physically motivated
law of differential rotation, essentially all of these works have assumed
that the rotation law is particularly simple and such that the specific
angular momentum $j:= h u_\phi$ is constant, with $h$ being the specific
enthalpy and $u_\phi$ the covariant azimuthal component of the
four-velocity. This law obviously satisfies the Rayleigh criterion for
local dynamical stability against axisymmetric perturbations, $dj/d\Omega
<0$, where $\Omega$ is the angular velocity.  More importantly, however,
it has the advantage of being analytically simple, with the angular
velocity decreasing \textit{monotonically} from the center of the star and
with the degree of differential rotation being regulated by a single
dimensionless parameter (normally referred to as $\tilde{A}$). Only
recently, Ref.~\cite{Galeazzi:2011} has made the first attempts to
generalize the space of equilibria by considering more general
prescriptions for the law of differential rotation, using however
analytical simplicity as the guideline.

Overall, all the studies carried out so far on equilibrium sequences of
stationary models agree that when using a $j-{\rm constant}$ law of
differential rotation, several different solutions are possible for a
given degree of differential rotation \cite{Ansorg2009, Gondek2016} and
that the maximum mass depends on both the degree of differential rotation
and on the type of solution, reaching values as large as four times that
of the maximum mass of nonrotating configurations $M_{_{\rm TOV}}$
\cite{Ansorg2009, Gondek2016}. This is to be contrasted with what happens
in the case of uniformly-rotating models, where the maximum mass has
recently been found to be only 20\% larger than $M_{_{\rm TOV}}$, quite
independently of the equation of state (EOS) \cite{Breu2016}.

While the $j-{\rm constant}$ law of differential rotation has been useful
so far to explore the equilibria of differentially rotating compact
stars, it is also hard to justify on physical grounds, in particular when
considering the results of numerical simulations of merging binary
neutron stars. A number of studies in this direction, in fact, have shown
that although the merger normally leads to the formation of a
hypermassive neutron star (HMNS), that is, a neutron star whose mass
exceeds the maximum mass of a uniformly rotating star, the law of
differential rotation is rather different~\cite{Shibata99d, Baiotti08,
  Anderson2007, Liu:2008xy, Bernuzzi2011, Rezzolla:2011,
  DePietri2016}. Hence, the need to determine a law of differential
rotation that, albeit not simpler, does reflect the hydrodynamical
equilibrium that is attained in HMNSs produced in binary neutron-star
mergers. Clearly, determining such a law of differential rotation is
important since it has direct impacts on the maximum mass these objects
can attain and hence on the threshold to black-hole formation under
astrophysically realistic conditions.

A first step in this direction has been taken in Refs.~\cite{Kastaun2014,
  Kastaun2016}, where the properties of the angular-velocity distribution
in the merged object have been analyzed in detail, although only for a
very limited number of binaries. In agreement with the previous
simulations of Ref.~\cite{Shibata06a}, these more recent works have found
that the angular-velocity profile of the HMNS shows a slowly rotating
core and an envelope that rotates at {angular frequencies} that scale
as $r^{-3/2}$, where $r$ is the radial coordinate in our coordinate
system.  This holds true both when the initial data is that of
irrotational binaries and when the binaries are artificially spun-up as
done in Ref.~\cite{Kastaun2013}.

Here, we adopt a more systematic approach and study the rotational
properties of the HMNSs produced by the merger of binary neutron stars
described with various EOSs and masses, and as computed via a large
number of numerical simulations in full general relativity. Besides
confirming the more specific results of Refs.~\cite{Shibata06a,
  Kastaun2014, Kastaun2016}, our most interesting finding is arguably
that the angular-velocity distribution shows only a modest dependence on
the EOS, thus exhibiting the traits of ``quasi-universality'' that have
been found in other aspects of compact stars, both when isolated
\cite{Yagi2013b, Pappas2014, Stein2014, Haskell2014, Doneva2014a,
  Chakrabarti2014, Breu2016} and in binary systems \cite{Yagi2013b,
  Maselli2013, Takami2014, Rezzolla2016}. More specifically, the
EOS-independent angular-velocity distributions we find are characterized
by an almost uniformly rotating core and a 
``disk'' with {angular frequencies $\Omega(r) \propto r^{-3/2}$}.

Having a ``disk'' in the outer regions of the HMNS is important for two
reasons, at least. Firstly, the disk surrounding the HMNS will accrete
onto the uniformly rotating core of the star only on a dissipative
timescale, thus not affecting its long-term stability. Secondly, once the
core of the HMNS eventually collapses to a rotating black hole, the
presence of a certain amount of mass on stable orbits will guarantee that
the black hole will not be ``naked'', but surrounded by a torus, as
expected when the collapse to a black hole is prompt
\cite{Baiotti08}. Both of these considerations are important within the
proto-magnetar model for short gamma-ray bursts~\citep{Zhang2001,
  Metzger2008, Bucciantini2012} and the subsequent extended X-ray
emission \cite{Rezzolla2014b} (see also \cite{Ciolfi2014} for a similar
model).

The paper is organized as follows: Section~\ref{sec:numrel} is dedicated
to a brief overview of the mathematical and numerical setup employed in
our simulations, while in Secs.~\ref{sec:hmb} and ~\ref{sec:lmb} we
illustrate the results obtained when modelling high- and low-mass
binaries, respectively. We focus, in particular, on the distributions of
rest-mass density and angular velocity, and illustrate our approach to
obtain time and azimuthally averaged profiles. Section~\ref{sec:tracers}
focuses on the use of tracer particles to disentangle the physically
meaningful results from the possible contamination of gauge effects,
while Sec.~\ref{sec:qub} discusses the ``quasi-universal'' features of
the angular-velocity profiles and how to correlate them with the
properties of the progenitor stars in the binary. Also discussed in
Sec.~\ref{sec:qub} are the amount of mass in the disk and the influence
of the thermal component of the EOS on the results presented. Finally, in
Sec.~\ref{sec:sum} we present a summary of our findings and the impact
they have on the stability of differentially rotating compact
stars. \lrn{Appendix \ref{sec:appendix_a} provides a discussion in terms of the
dynamics of tracers on the conservation of the Bernoulli constant in the
quasi equilibrium of the HMNS. Appendix \ref{sec:appendix_b} provides a
discussion of the effects of resolution and symmetries on the lifetime and
evolution of the HMNS.}

\noindent Hereafter, we will use a spacelike signature $(-,+,+,+)$ and a
system of units in which $c=G=\Msun=1$ unless stated differently.

%%%%%%%%%%%%%%%%%%%%%%%%%%%%%%%%%%%%%%%%%%%%%%%%%%%%%%%%%%%%%%%%%%
\section{General Framework}
\label{sec:numrel}
%%%%%%%%%%%%%%%%%%%%%%%%%%%%%%%%%%%%%%%%%%%%%%%%%%%%%%%%%%%%%%%%%%

\subsection{Mathematical and numerical setup}

The mathematical and numerical setup used for the simulations reported
here is the same discussed in Refs. \cite{Takami2014, Takami2015} and
presented in greater detail in other papers \cite{Baiotti08,
  Baiotti:2009gk, Baiotti:2010ka}. For completeness we review here only
the basic aspects, referring the interested reader to the papers above
for additional information.

Our simulations are performed in general relativity using the
fourth-order finite-differencing code \texttt{McLachlan}
\cite{Brown:2008sb, Loffler:2011ay}, which is part of the publicly
available \texttt{Einstein~Toolkit}~\cite{einsteintoolkitweb}. This code
solves a conformal traceless formulation of the Einstein equations
\cite{Nakamura87, Shibata95, Baumgarte99}, with a ``$1+\log$'' slicing
condition and a ``Gamma-driver'' shift condition
\citep{Alcubierre02a,Pollney:2007ss}. Correspondingly, the evolution of
the general-relativistic hydrodynamics equations is done using the
finite-volume code \texttt{Whisky} \citep{Baiotti04}, which has been
extensively tested in simulations involving the inspiral and merger of
binary neutron stars \cite{Baiotti08, Baiotti:2009gk, Rezzolla:2010,
  Baiotti:2010}. The hydrodynamics equations, expressing the conservation
of energy, momentum and rest mass \cite{Rezzolla_book:2013}, are cast in
the conservative Valencia formulation~\cite{Font08}. Their numerical
solution is obtained employing the Harten-Lax-van Leer-Einfeldt
\cite{Harten83} approximate Riemann solver \cite{Harten83} in conjunction
with the Piecewise Parabolic Method \cite{Colella84} for the
reconstruction of the evolved variables. For the time integration of the
coupled set of hydrodynamic and Einstein equations we use the Method of
Lines with an explicit fourth-order Runge-Kutta method. Our simulations
use a CFL number of $0.35$ to compute the timestep.

An adaptive mesh refinement (AMR) approach based on the \texttt{Carpet}
mesh-refinement driver~\cite{Schnetter-etal-03b} is used to both increase
resolution and extend the spatial domain, placing the outer boundary as
close as possible to the wave zone. The grid hierarchy consists of six
refinement levels with a grid resolution varying from $\ensuremath{\Delta
  h_5 = 0.15\,M_\odot}$ (\ie $\simeq 221\,{\rm m}$) for the finest level
to $\ensuremath{\Delta h_0 = 4.8\,M_\odot}$ (\ie $\simeq 7.1\km$) for the
coarsest level, whose outer boundary is at $514\,M_\odot$ (\ie $\simeq
759\km$). To reduce computational costs, a reflection symmetry across the
$z=0$ plane and a $\pi$-symmetry condition across the $x=0$ plane has
been adopted \lrn{(see Appendix \ref{sec:appendix_b} for a discussion of
  the influence on the results of this resolution and of the symmetry
  assumed)}. The initial configuration for the quasi-equilibrium
irrotational binary neutron stars has been generated with the use of the
\texttt{LORENE}-code~\citep{Gourgoulhon-etal-2000:2ns-initial-data} and
an initial coordinate separation of the stellar centers of
$45\,\mathrm{km}$ has been used for all binaries. For each EOS, we have
considered equal-mass binaries where each star has initial gravitational
masses that are either $M=1.25 \Msun$ (low-mass binaries; \cf
Sec. \ref{sec:lmb}) or $1.35 \Msun$ (high-mass binaries; \cf
Sec. \ref{sec:hmb}) at infinite separation.

A quantity that is particularly important in our analysis is the angular
velocity, $\Omega$, which is defined as the amount of coordinate rotation
\begin{equation}
\Omega := \frac{d\phi}{dt} = \frac{d x^\phi}{dt} =
\frac{u^\phi}{u^t} \,,
\label{eq:omega1}
\end{equation}
where $u^\phi$ and $u^t$ are components of the four-velocity vector
$u^\mu$.  The corresponding three-velocity as measured by the same
Eulerian observer is then defined as
\begin{equation}
v^i :=  \frac{\gamma^i_\mu u^\mu}{-n_\mu u^\mu} = 
\frac{1}{\alpha} \left(\frac{u^i}{u^t} + \beta^i \right)\,, 
\label{eq:v1}
\end{equation}
where $\alpha$ is the lapse function, $\beta^{i}$ is the shift vector,
$n^\mu$ is the unit timelike vector normal to a constant $t$
hypersurface, and $\gamma_{ij}$ is the three-dimensional metric. With the
above definition the angular velocity within the $3+1$ split can be
expressed as
\begin{equation}
\Omega = \alpha~ v^\phi~ - ~\beta^\phi \,,
\label{eq:omega2}
\end{equation}
where
\begin{align}
v^\phi=\frac{x v^y - y v^x}{x^2+y^2+z^2}\,,\\
\beta^\phi=\frac{x \beta^y - y \beta^x}{x^2+y^2+z^2}\,,
\end{align}
are the three-velocity and shift vector components as computed from the
Cartesian grid with coordinates $(x,y,z)$. Written in this form, we can
interpret $\Omega$ as consisting of a lapse-corrected part of the
$\phi$-component of the three-velocity, minus a frame-dragging term
provided by the $\phi$-component of the shift vector.

%-------------------------------------------------------
\subsection{Microphysical matter treatment}
\label{sec:eos}
%-------------------------------------------------------

To close the set of evolution equations an EOS is needed and which
provides a relation among the thermodynamical properties of the
neutron-star matter. A general EOS describes the pressure as a function
of the rest-mass density, particle composition and temperature. However,
for certain parts of the merger, simplified versions can be used. In
particular, during the inspiral, a ``cold'', \ie temperature independent,
EOS is sufficient to represent the state of the neutron-star matter prior
to merger. After contact, when the HMNS is formed, large shocks will
increase the temperature and to account for this additional heating a
full temperature dependent nuclear-physics EOS is required. As the number
of such EOSs is unfortunately still limited, we include thermal effects
by adding an ideal-fluid component that accounts for the shock
heating. The pressure $p$ and the specific internal energy $\epsilon$ are
therefore composed of a cold nuclear-physics part and of a ``thermal''
ideal-fluid component\footnote{This class of EOSs is referred in the
  literature as the so-called ``hybrid EOS'' \cite{Rezzolla_book:2013}
  and should not be meant to indicate that the star is composed of a
  hybrid hadron-quark matter present in hybrid stars.} \cite{Janka93}
\begin{equation}
\label{EOS:full_a}
 p = p_\mathrm{c} + p_\mathrm{th}\,, \qquad
\epsilon = \epsilon_\mathrm{c} + \epsilon_\mathrm{th} \,,
\end{equation}
where $p$ and $\epsilon$ are the pressure and specific internal energy,
respectively. We model the cold part $p_\mathrm{c}, \epsilon_\mathrm{c}$
with five different nuclear-physics EOSs. Two of such EOSs, namely
APR4~\cite{Akmal1998a} and SLy~\cite{Douchin01}, belong to the class of
variational-method EOSs and the underlying particle composition within
these models consists mainly of neutrons with little admixtures of
protons, electrons and muons. Additionally, two more EOSs, \ie
GNH3~\cite{Glendenning1985} and H4~\cite{GlendenningMoszkowski91}, are
built using relativistic mean-field models which include, above a certain
rest-mass density, hyperonic particles.

\begin{table*}
\begin{footnotesize}
\begin{tabular}{l|c|c|c|c|c|c|c|c|c|c|c}
\hline
\hline
model 
& EOS             & ${M}$          & ${R}$ 
& $M_\mathrm{ADM}$  & ${M_\mathrm{b}}$ & ${M}/{R}$  
& $f_\mathrm{orb}$  & $J$            & ${I}/{M}^3$ 
& ${k}_{2}$        & $\lambda/{M}^5$  \\
&                 & $[\Msun]$          & $[\mathrm{km}]$    
& $[\Msun]$       & $[\Msun]$          &
& $[\mathrm{Hz}]$ & $[\Msun^2]$        &
&                 &\\
\hline
\texttt{GNH3-M125} & GNH3 & $1.250$ & $13.817$ & $2.4780$ & $1.3464$ & $0.13358$ & $273.29$ & $6.4067$ & $18.890$ & $0.11753$  & $1842.4$  \\
\texttt{GNH3-M135} & GNH3 & $1.350$ & $13.777$ & $2.6746$ & $1.4641$ & $0.14468$ & $281.58$ & $7.2766$ & $16.450$ & $0.10841$  & $1139.9$  \\
\hline
\texttt{H4-M125}   & H4   & $1.250$ & $13.533$ & $2.4780$ & $1.3506$ & $0.13638$ & $273.25$ & $6.4058$ & $18.610$ & $0.12361$  & $1746.5$  \\
\texttt{H4-M135}   & H4   & $1.350$ & $13.550$ & $2.6746$ & $1.4687$ & $0.14711$ & $281.61$ & $7.2770$ & $16.344$ & $0.11483$  & $1111.1$  \\
\hline
\texttt{ALF2-M125} & ALF2 & $1.250$ & $12.276$ & $2.4779$ & $1.3672$ & $0.15034$ & $273.16$ & $6.4014$ & $16.455$ & $0.13049$  & $1132.6$  \\  
\texttt{ALF2-M135} & ALF2 & $1.350$ & $12.353$ & $2.6746$ & $1.4877$ & $0.16136$ & $281.42$ & $7.2708$ & $14.581$ & $0.12037$  & $733.63$  \\
\hline
\texttt{SLy-M125}  & SLy  & $1.250$ & $11.469$ & $2.4779$ & $1.3720$ & $0.16092$ & $273.04$ & $6.3977$ & $14.000$ & $0.10266$  & $634.27$  \\
\texttt{SLy-M135}  & SLy  & $1.350$ & $11.465$ & $2.6745$ & $1.4946$ & $0.17386$ & $281.34$ & $7.2663$ & $12.309$ & $0.092993$ & $390.29$  \\
\hline
\texttt{APR4-M125} & APR4 & $1.250$ & $11.052$ & $2.4779$ & $1.3783$ & $0.16700$ & $273.05$ & $6.3973$ & $13.226$ & $0.099787$ & $512.14$  \\
\texttt{APR4-M135} & APR4 & $1.350$ & $11.079$ & $2.6746$ & $1.5020$ & $0.17992$ & $281.37$ & $7.2665$ & $11.720$ & $0.090990$ & $321.78$  \\
\hline
\texttt{LS220-M132}& LS220 & $1.319$ & $12.775$ & $2.6127$ & $1.4360$ & $0.15108$ & $278.68$ & $6.9891$ & $15.113$ & $0.099289$ & $840.93$  \\
\texttt{LS220-M135}& LS220 & $1.350$ & $12.750$ & $2.6740$ & $1.4733$ & $0.15638$ & $281.29$ & $7.2656$ & $14.112$ & $0.096575$ & $688.26$  \\ %%
\hline
\hline
\end{tabular}
\caption{All binaries evolved and their properties. The various columns
  denote the gravitational mass $M$ of the binary components at infinite
  separation, the corresponding radius $R$, the ADM mass $M_\mathrm{ADM}$
  of the binary system at the initial separation, the baryon mass
  ${M}_\mathrm{b}$, the compactness ${\mathcal C}:= {M}/{R}$, the orbital
  frequency $f_\mathrm{orb}$ at the initial separation, the total angular
  momentum $J$ at the initial separation, the dimensionless moment of
  inertia ${I}/{M}^3$ at infinite separation, the $\ell=2$ dimensionless
  tidal Love number ${k}_2$ at infinite separation, and the dimensionless
  tidal deformability $\lambda/{M}^5$ defined by $\lambda:= 2 {k}_2
  {R}^5/3$.
\label{tab:models}}
\end{footnotesize}
\end{table*}

In contrast, the fifth EOS, namely ALF2~\cite{Alford2005}, is more a
model for hybrid stars than for neutron stars because it implements a
phase transition to color-flavor-locked quark matter. Within this model,
the hadronic particles begin to deconfine to quark matter above a certain
transition rest-mass density $\rho_{{\rm trans}}=3\,\rho_{\rm nuc}$,
where $\rho_{\rm nuc}:=2.705\times10^{14}$\, g/{cm$^3$} is the
nuclear-matter rest-mass density. Assuming a moderate surface tension of
the quark matter droplets, a phase transition is implemented by using a
Gibbs-construction. As charge neutrality is only globally conserved
within this construction, a mixed-matter phase exists in the rest-mass 
density range $3\,\rho_{\rm nuc} \leq \rho \leq 7.8\,\rho_{\rm nuc}$.

We note that all of the EOSs used in our calculations satisfy the current
observational constraint on the observed maximum mass in neutron stars,
\ie $2.01\pm0.04M_\odot$ \cite{Antoniadis2013}. Instead of using the data
tables of the various EOSs, we have found it more suitable to convert
them to piecewise polytropes~\cite{Read:2009a}. Each EOS has been
parametrized, based on specifying its stiffness in three rest-mass
density intervals $i=2,3,4$, measured by the adiabatic index
$\Gamma_i=d\,\log\bar{p}_i/d\,\log\bar{\rho}_i$. Additionally, a unique
polytrope with $\Gamma_1=1.357$ has been added for all of the used EOS to
account for the star's low rest-mass density region $\rho \leq \rho_{\rm
  nuc}$. 

The ``cold'' nuclear-physics contribution to each EOS is obtained after
expressing the pressure and specific internal energy
$\epsilon_\mathrm{c}$ in the rest-mass density range $\rho_{i-1} \leq
\rho < \rho_i$ as (for details see \cite{Bauswein2012a, Bauswein2012,
  Takami:2014, Takami2015})
\begin{equation}
\label{EOS:cold_a}
p_\mathrm{c} = K_i \rho^{\Gamma_i}\,,\qquad
\epsilon_\mathrm{c} = \epsilon_i + K_i
\frac{\rho^{\Gamma_i-1}}{\Gamma_i -1} \,. 
\end{equation}
For an overall consistency, the rest-mass density ranges used for the
piecewise polytropes have been chosen to be the same for the different
EOSs ($\rho_2=5.012 \times 10^{14}\gcm$ and $\rho_3 = 10^{15}\gcm$). The
transition densities $\rho_1$ to the low rest-mass density polytrope and
the EOS-dependent adiabatic indexes $\Gamma_i$ are summarized in Table 2
of \cite{Takami:2014}. Due to the implementation of the hadron-quark
phase transition, the ALF2 EOS has the largest softening at the rest-mass
density boundary $\rho_2$ ($\Gamma_1=4.070$ and
$\Gamma_2=2.411$). Finally, the ``thermal'' part of the EOS is given by
\begin{equation}
\label{EOS:hot}
p_\mathrm{th} = \rho \epsilon_\mathrm{th}
\left(\Gamma_\mathrm{th} -1 \right)\,,\qquad
\epsilon_\mathrm{th}  = \epsilon - \epsilon_\mathrm{c} \,. 
\end{equation}
where the last equality in \eqref{EOS:hot} is really a definition, since
$\epsilon$ refers to the computed value of the specific internal energy.
In all of the simulations reported hereafter we use
$\Gamma_\mathrm{th}=2.0$ (see \cite{Takami:2014, Takami2015} for an
analysis of the effect of different $\Gamma_\mathrm{th}$ and
Sec. \ref{sec:itc} for a discussion of the impact of $\Gamma_\mathrm{th}$
on the results presented here). Detailed information on all the binaries
and their properties is collected in Table \ref{tab:models}.

Finally, we note that in order to verify our hybrid-EOS approach and
numerical setup and contrast the results with an alternative one, an
additional simulation has been performed using the ``hot'', \ie
  temperature dependent, Lattimer-Swesty (LS220) EOS
\cite{Lattimer91}. For this simulation the hydrodynamic equations are
solved employing the \texttt{WhiskyTHC} code \cite{Radice2013b,
  Radice2013c} and the BSSNOK formulation of the Einstein equations
\cite{Alic:2011a}.

%%%%%%%%%%%%%%%%%%%%%%%%%%%%%%%%%%%%%%%%%%%%%%%%%%%%%%%%%%%%%%%%%%
\section{High-mass binaries}
\label{sec:hmb}
%%%%%%%%%%%%%%%%%%%%%%%%%%%%%%%%%%%%%%%%%%%%%%%%%%%%%%%%%%%%%%%%%%

The main results concerning the evolution of the characteristic
properties of the HMNSs are presented in this and the following
section. To explain the procedure of how the HMNS properties and
rotational profiles are calculated in our approach, we start by
describing in this section the dynamics of a representative ``high-mass
binary'' focusing on the \texttt{ALF2-M135} run, that is, a binary
described by the ALF2 EOS and with total gravitational mass of
$2\times 1.35\,M_{\odot}$; in the following section, we describe instead the
\texttt{ALF2-M125} run as an illustrative case of ``low-mass
simulations''.

\begin{figure}
\begin{center}
\includegraphics[width=\columnwidth]{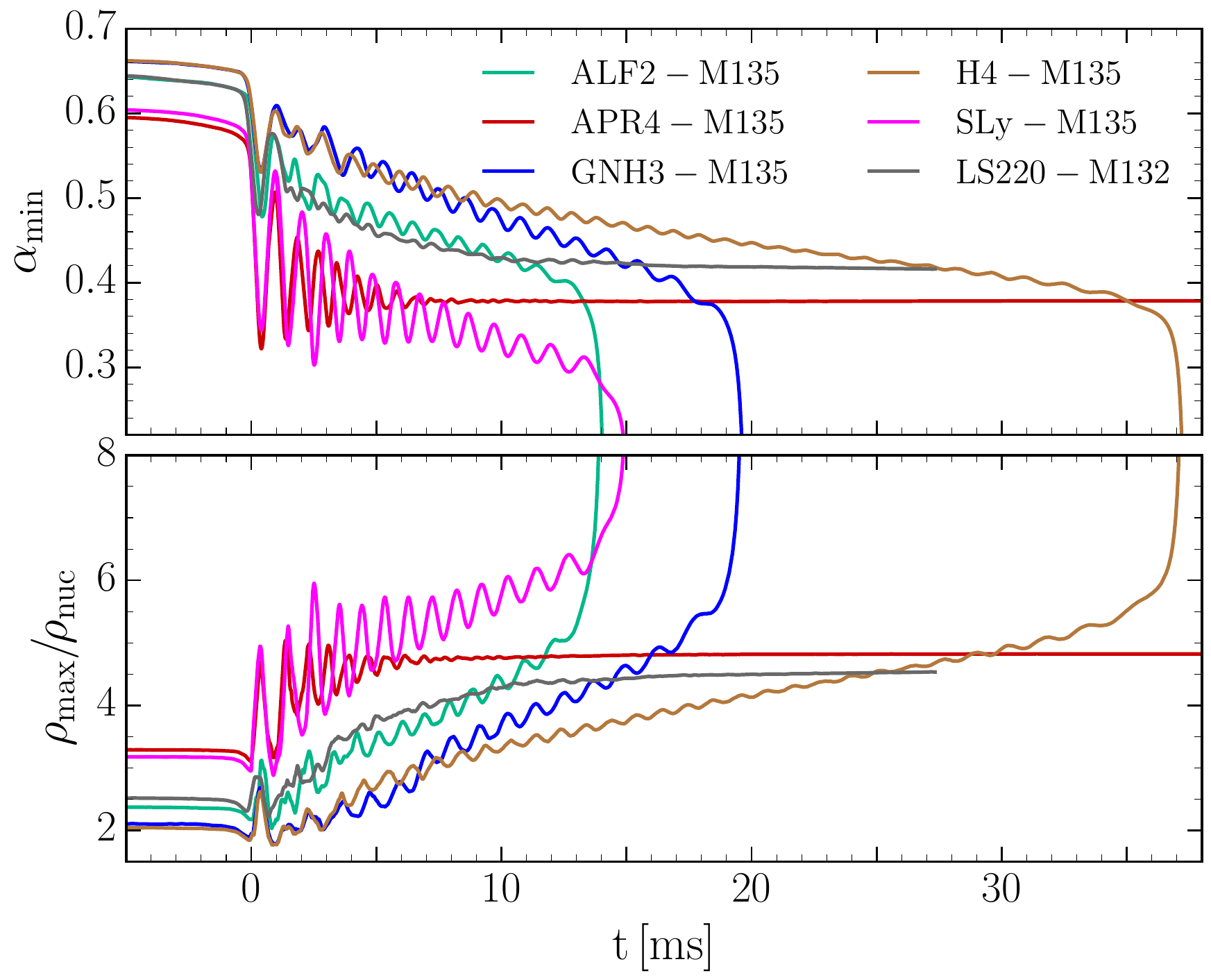}
\caption{Minimum value of the lapse function $\alpha_{{\rm {min}}}$
  (upper panel) and maximum of the rest-mass density $\rho_{{\rm {max}}}$
  in units of the nuclear-matter rest-mass density $\rho_{\rm nuc}$
  (lower panel) versus time in milliseconds after the merger for the
  high-mass simulations. All models collapse to a black hole except
  for the APR4 and LS220 EOSs.}\label{fig:alphatimeH}
\end{center}
\end{figure}

We recall that with the exception of the APR4 and LS220 EOSs, four of the
six binaries with high masses ($M=1.35 M_\odot$) collapse to a black hole
within the simulated timescale, while for the low-mass cases ($M=1.25
M_\odot$) none of the simulations show a gravitational collapse within
our simulation time domain (see Figs. \ref{fig:alphatimeH} and
\ref{fig:alphatimeL}).

Before the merger, the maximum (central) value of the rest-mass density
$\rho_{{\rm {max}}}$ is essentially constant in time and the minimum
(central) value of the lapse function $\alpha$ decreases only slightly
(see Fig.~\ref{fig:alphatimeH} for $t<0$, where we define as $t=0 :=
t_{{\rm {M}}}$ the time of merger or, equivalently, the time of the first
maximum in the gravitational-wave amplitude). The differences of the
values of $\rho_{{\rm {max}}}$ and $\alpha_{{\rm {min}}}$ before merger
are due to the softness/stiffness properties of the underlying EOSs,
which determine the HMNS individual compactness. Although the HMNS is
formed at the merger time, the rest-mass density profile at $t_{{\rm
    {M}}}=0$ still has two distinct maxima that correspond to the tidally
deformed individual stars of the late inspiral phase. In the transient
post-merger phase (\ie $t\in[0,4]\ms$) irregular and strong
fluctuations of $\rho_{{\rm {max}}}$ and $\alpha_{{\rm {min}}}$ occur,
which are due to the violent and shock-dominated dynamics right after the
merger. Within this early post-merger stage, the rest-mass density
profiles of the HMNSs have two distinct maxima (the so-called
``double-core'' structure), which indicates that $\rho_{{\rm {max}}}$ and
$\alpha_{{\rm {min}}}$ no longer correspond to the central HMNS values,
respectively (see Ref. \cite{Takami:2014} for a simple mechanical toy
model that describe this phase of the post-merger).

At later times (\ie $t \gtrsim 4\ms$) the two maxima merge to one single
maximum at the HMNS's center. This feature holds for all performed
simulations with different EOSs and masses, and is in accordance with
many other works (\eg \cite{Shibata:2003ga,Baiotti08}). Within this
``post-transient'' phase, $\rho_{{\rm {max}}}$ ($\alpha_{{\rm {min}}}$)
show a quite regular oscillating behaviour with an average increasing
(decreasing) value. Additionally, it is possible to note a continuous
decrease in the oscillation frequency, which signals the approaching of
the ``zero-frequency'' limit and hence the quasi-radial stability limit
to gravitational collapse \cite{Takami:2011}. The lifetime of the HMNSs
is different for the various EOSs and the collapse to the final black
hole can be easily seen in Fig.~\ref{fig:alphatimeH} as a sudden singular
increase in $\rho_{{\rm {max}}}$ (sudden decrease in $\alpha_{{\rm
    {min}}}$).

\begin{figure}%[t]
\begin{center}
\includegraphics[width=\columnwidth]{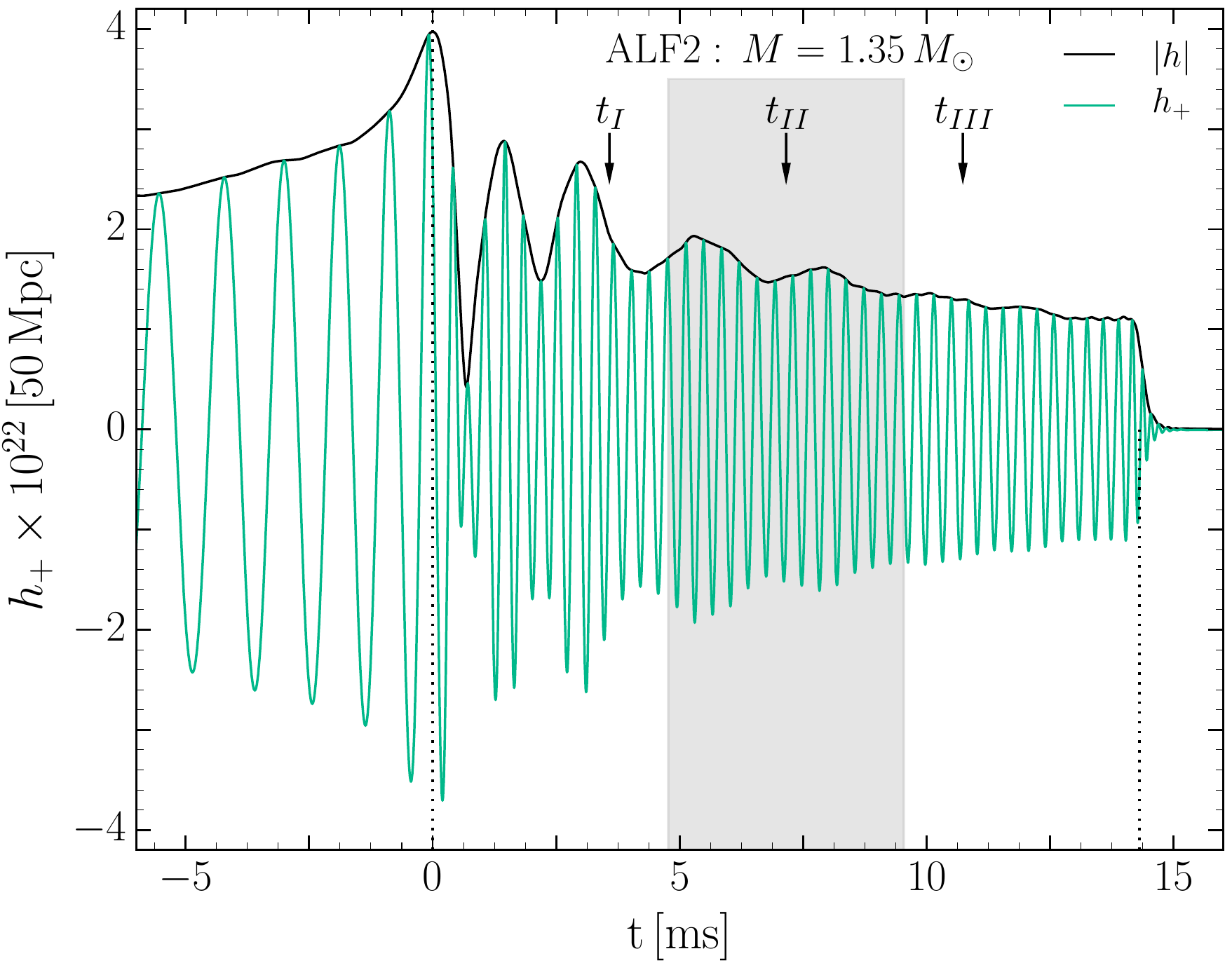}
\caption{Gravitational-wave amplitude $|h|$ (black line) and strain
  amplitude in the $+$ polarisation $h_+$ (green line) for the
  \texttt{ALF2-M135} binary at a distance of 50 Mpc. Shaded in gray is
  the portion where a time average is performed, while the arrows
  indicate the times when representative distributions of the rest-mass
  density and angular velocity are shown in
  Fig.~\ref{fig:ALF2-M135-rho}. Finally, the dotted vertical lines mark
  the time of merger and the first detection of an apparent horizon.}
\label{fig:GW-ALF2-M135}
\end{center}
\end{figure}

%-------------------------------------------------------
\subsection{Density evolution and gravitational-wave emission}
%-------------------------------------------------------

Before merger, the two individual stars of the \texttt{ALF2-M135} binary
have a central rest-mass density $\rho_{{\rm max}} \simeq 2.17\,\rho_{\rm
  nuc}$ that is below the onset of a hadron-quark phase transition, $\rho
< 3\,\rho_{\rm nuc}$. Hence, all of the matter inside the HMNS at merger
time is mainly composed of neutrons with little admixtures of protons and
electrons \cite{Alford2005}. Within the early transient post-merger
phase, $\rho_{{\rm {max}}}$ reaches values above $3\,\rho_{\rm nuc}$, but
due to the strongly oscillating nature of the double core structure, the
deconfined mixed phase does not remain in the HMNS and confines again to
hadronic matter. Nevertheless, a considerable amount of hadronic matter
deconfines in this transient post-merger phase to quark matter and in
less than 0.5 ms it confines again to hadronic matter. We recall that
under the ``strange matter hypothesis'', the strange-quark phase is the
true ground state of matter and the whole neutron star would then
transform into a pure-quark star after exceeding a certain deconfinement
barrier \cite{Haensel07, Drago2015, Bombaci2016, Lugones2016}. During
such a process a significant amount of energy would be released in the
form of neutrinos and gamma-rays \cite{Drago2016}. The quark-matter
nucleation process, which takes place in the highly dynamical interior
region of the HMNS, depends on various poorly known factors (\eg the
surface tension). In the ALF2 EOS, the strange-matter hypothesis has not
been adopted and as a consequence, stable hybrid stars can be formed,
having a inner core of deconfined strange matter.

\begin{figure*}
\begin{center}
  \includegraphics[width=2.0\columnwidth]{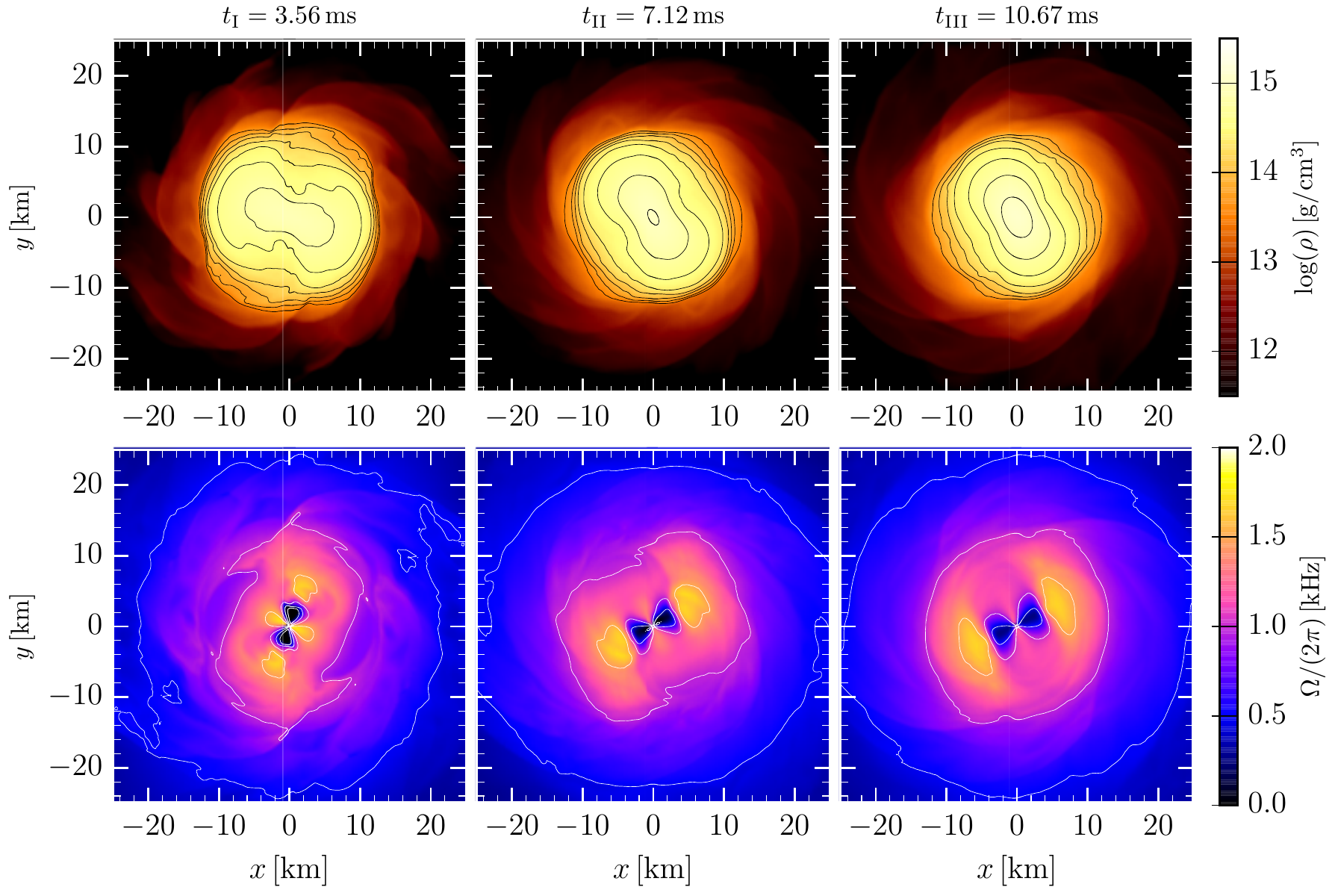}
  \caption{Distributions of the rest-mass density (upper row, log scale)
    and of the fluid angular velocity (lower row) in the $(x,y)$ plane
    for the \texttt{ALF2-M135} binary at three different post-merger
    times as indicated in Fig.~\ref{fig:GW-ALF2-M135}. The isocontours
    have been drawn at $\log(\rho) = 13.6 + 0.2n$ (upper row) and $\Omega
    = \{0,0.5,1.0,1.5,2.0\} \, {\rm kHz}$ (lower row), $n \in
    \mathbb{N}$.}
\label{fig:ALF2-M135-rho}
\end{center}
\end{figure*}

In some other EOS-models of this kind, \eg \cite{Glendenning2000, 
Mishustin2003, Macher2005, Alford2015, Zacchi2016}, there exist however 
parameter ranges where ``twin stars'' solutions are present \cite{Haensel07} 
and transitions between these twins could likewise produce neutrino and 
even short gamma-ray bursts, which we do not account for in these
simulations. Approximately 5 ms after merger, $\rho_{{\rm {max}}}$ is
permanently above the phase-transition threshold and the HMNS contains an
inner region of deconfined mixed-phase matter. A pure-quark phase appears
only during the short collapse phase to black hole and finally this free
quark matter will be macroscopically deconfined by the growth of the
event horizon inside the collapsing HMNS.

For the gravitational-wave amplitudes $h_+$ and $h_\times$ we consider only the
$(\ell,m)=(2,2)$ mode, which has been found to be the most dominant (for
details see \cite{Takami:2014,
  Reisswig:2011}). Figure~\ref{fig:GW-ALF2-M135} shows the
gravitational-wave amplitude $|h| := ( {h^2_+} + {h^2_{\times}} )^{1/2}$
and $h_+$ at a distance of 50 Mpc as a function of time. The absolute
maximum of $|h|$ corresponds to the time of merger for all of the
different simulation runs. The last peak of $h_+$ corresponds
approximately to the time when the black hole is formed (\ie$t_{{\rm
    {BH}}} = 14.16\ms$), which we have defined as the time when the
apparent horizon is first detected. In order to compare the structural
properties of the different HMNSs in the post-merger phase, we will later
define a time-averaging procedure of the rotation profiles. This
averaging time interval is already shown in Fig.~\ref{fig:GW-ALF2-M135}
as a gray region.

The upper row of panels in Fig.~\ref{fig:ALF2-M135-rho} shows the
evolution of the HMNS rest-mass density during the post-merger phase. The
three different snapshots have been taken at $t=1/4$, $1/2$ and
$3/4\,t_{{\rm {BH}}}$. The left panel, in particular, which visualizes
the rest-mass density distribution at $t_{{\rm {I}}} = t_{{\rm {BH}}}/4
\approx 3.6\ms$, shows that the overall rest-mass density of the HMNS is
much higher than the rest-mass density at merger time. The double-core
structure in the inner area of the HMNS is right on the verge of merging
to a single core and the maximum rest-mass density reached, $\rho_{{\rm
    {max}}} \simeq 3.1\rho_{\rm nuc}$, is slightly above the onset of the
underlying hadron-quark phase transition. In the central and right upper
panels of Fig.~\ref{fig:ALF2-M135-rho} ($t_{_{\rm II}} = t_{{\rm {BH}}}/2
\approx 7.1\ms$ and $t_{_{\rm III}} = 3\,t_{{\rm {BH}}}/4 \approx 10.7$
ms) the double-core structure is no longer present. The value of the
central rest-mass density maximum $\rho_c \simeq 3.8\,\rho_{\rm nuc}$ at
$t_{_{\rm II}}$ is clearly above the onset of the underlying hadron-quark
phase transition. Since the ALF2 EOS uses a Gibbs construction for the
modelling of the phase transition (see Section \ref{sec:numrel}), the
inner core ($r\lesssim 3\km$) of the HMNS contains a crystalline
structure of mixed phase matter, which might have additional effects on
the evolution of the HMNS.

%-------------------------------------------------------
\subsection{Angular-velocity evolution: 2D slices}
\label{sec:omega2D}
%-------------------------------------------------------

\begin{figure*}
\begin{center}
\includegraphics[width=2\columnwidth]{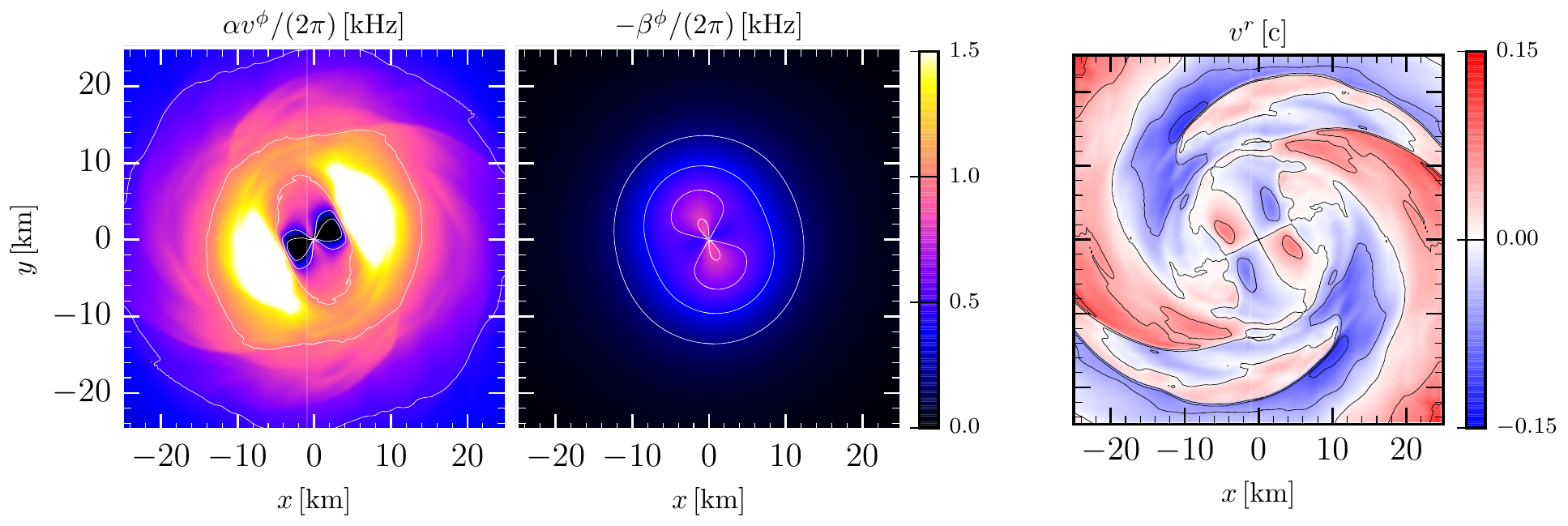}
\caption{Distributions on the $(x,y)$ plane for the
  \texttt{ALF2-M135} binary at $t_{_{\rm III}}$ of the two contributions
  to the angular velocity $\Omega$. The left and central panels refer to
  the quantities $\alpha v^{\phi}$ and $-\beta^{\phi}$, respectively [\cf
    Eq. \eqref{eq:omega2}]. The right panel shows instead the
  distribution of the radial component of the fluid three-velocity
  $v^r$.}\label{fig:ALF2-M1350-beta}
\end{center}
\end{figure*}

In the following, and to contrast the description made above of the
rest-mass density, we will concentrate on the evolution in the
post-merger phase of the angular velocity $\Omega$ [see
  Eq.~(\ref{eq:omega2})], which plays a particularly important role in
our analysis. The lower panels of Fig.~\ref{fig:ALF2-M135-rho} display
the distribution of $\Omega$ in the $(x,y)$ plane at the same times shown
for the rest-mass density in the upper panels. In the first two
milliseconds after the merger the time variation of $\Omega$ is very
rapid, with two inner maxima placed between the double-core rest-mass
density maxima (left lower panel). In the intermediate part of the
post-merger (\ie for $t \in [3 \,{\rm ms}, t_{{\rm {BH}}}]$), $\Omega$
has a roughly time independent global structure in a frame corotating at
half the frequency of the gravitational-wave emission, $\Omega_{_{\rm
    GW}}$ (see Sec.~\ref{sec:tracers}); this structure remains stationary
for several milliseconds (right lower panel). Approximately two
milliseconds before black hole formation, $\Omega$ largely increases in
the center of the HMNS (not shown in Fig.~\ref{fig:ALF2-M135-rho}.)

Note that also the angular-velocity distribution exhibits a clear $m=2$
distribution, where the two maxima rotate with the same frequency $\simeq
1.4\,{\rm kHz}$ around the center of the HMNS as the non-axisymmetric
$m=2$ perturbations of the rest-mass density (see upper middle panel of
Fig.~\ref{fig:ALF2-M135-rho}). Both maxima appear at a radial distance of
$\simeq 6\km$, which is still clearly within the star's high rest-mass
density range, but are outside the region where the hadron-quark phase
transition appears. The maxima in $\Omega$ are also accompanied by two
minima in the inner regions of the HMNS, where the angular velocity can
even become negative, \ie with the minima counter-rotating relative to
the outer layers of the HMNS.  \lrn{The largest gradients in the
  $\Omega$-profile take place at $\simeq 3\km$ from the center.}
Although both $\rho$ and $\Omega$ have marked $m=2$ distributions, it is
also clear that there is an evident phase offset of $\simeq 90$ degrees
between them. This feature is very robust and is present in all of the
binaries we have simulated. However, to the best of our knowledge, this
has not been reported before in the literature. \lrn{This phase offset
  can be explained rather simply in terms of an extension of Bernoulli's
  theorem for which areas of low (pressure) rest-mass density are
  accompanied by regions of large velocity; this is in essence what the
  upper and lower panels of Fig. \ref{fig:ALF2-M135-rho} express. A more
  detailed discussion of this point will be presented in Appendix
  \ref{sec:appendix_a}, when analysing the conservation of the Bernoulli
  constant associated to representative tracers.}

A question that arises when investigating the spatial properties of the
angular velocity $\Omega$ is whether such properties are physical and not
just a gauge artefact given that $\Omega$ is, after all, a
gauge-dependent quantity. Although the influence of gauge deformations
has already been assessed to be small in a study performed with very
similar gauges and physical conditions \cite{,Kastaun2016}, we have
performed a number of additional investigations to rule this out.
Firstly, we have investigated the evolution of the relevant components of
the spatial three-metric (namely, $\gamma_{rr}, \gamma_{\phi\phi}$ and
$\gamma_{r\phi}$) and did not find a corresponding structure which could
have produced the properties discussed above for the spatial distribution
of $\Omega$. Secondly, we have investigated the properties of the various
quantities that contribute to the angular velocity, namely, the frame
dragging provided by the shift component $-\beta^\phi$ and the azimuthal
fluid velocity $\alpha v^{\phi}$ [\cf Eq. \eqref{eq:omega2}], together
with the radial component of the three-velocity $v^r$.

Figure~\ref{fig:ALF2-M1350-beta} shows the equatorial structure of these
three quantities at time $t_{_{\rm III}}$. The left panel, in particular,
displays $\alpha v^{\phi}$, and has almost the same global structure as
that shown by $\Omega$ (see lower right panel of
Fig.~\ref{fig:ALF2-M135-rho}), with the only (obvious) difference that
the maximum and minimum amplitudes are smaller. The central panel of
Fig.~\ref{fig:ALF2-M1350-beta} shows the shift component $\beta^\phi$
with the same colorcode, indicating that it not only has a similar
spatial structure to the rest-mass density, but it is also considerably
smaller, becoming essentially zero in the outer layers of the HMNS (\ie
for $r \gtrsim 15\km$). Both of these facts indicate that the influence
of gauge quantities on the values of $\Omega$ cannot be responsible for
the 90-degrees phase shift, which has instead a rather intuitive
explanation given above.

In addition, the right panel of Fig.~\ref{fig:ALF2-M1350-beta} shows the
radial component of the three-velocity $v^r$, with red regions indicating
fluid cells with outward radial motion, while blue regions refer to fluid
moving inward. Note that at the outer parts of the HMNS (\ie for $r
\gtrsim 15\km$) the flow is mostly outwards along the two dense spiral
arms. These will feed the matter ejected dynamically that will eventually
lead to the production of heavy elements \cite{Wanajo2014,Radice2016}. On
the other hand, the distribution of the radial velocity in the inner
parts of the HMNS (\ie for $r\lesssim 10\km$) shows a clear quadrupolar
structure produced by the propagation of the $m=2$ rest-mass density
perturbation. To clarify the properties of this structure it is
sufficient to imagine an $\ell=m=2$ tidal wave moving along the surface
of an otherwise spherical star. The local velocity will be a $\ell=2,
m=4$ succession of positive and negative radial velocities as the tidal
wave sweeps through the surface. Additional considerations along these
lines will be made when discussing the motion of tracer particles in
Sec. \ref{sec:tracers}.

In summary, all the analyses discussed so far indicate that the $m=2$
distribution of the angular velocity presented in
Fig. \ref{fig:ALF2-M135-rho} is not contaminated by gauge effects and it
can be interpreted in terms of the physical manifestation of the
Bernoulli theorem.

Before concluding this section we should remark that all the results
presented above have been focused on the equatorial plane; however, the
extensions to planes at nonzero elevation is straightforward. More
specifically, we have found that the structure of the rest-mass density
and rotation profiles do not change significantly for \lrn{$z < 8\km$},
although the $m=2$ deformation tends to be less marked. Moving further
away from the equatorial plane, \lrn{the angular velocity becomes almost
  axisymmetric, being small in the core of the HMNS, but growing to
  larger values in the outer layers, \ie for $z > 11\km$. At these
  heights,} the rest-mass density has decreased considerably and the HMNS
smoothly blends in with the outward-moving wind of the dynamically
ejected matter and which will represent an important element of pollution
of the interstellar medium \cite{Murguia-Berthier2014,
  Murguia-Berthier2016}.

\begin{figure}%[b]
\begin{center}
\includegraphics[width=\columnwidth]{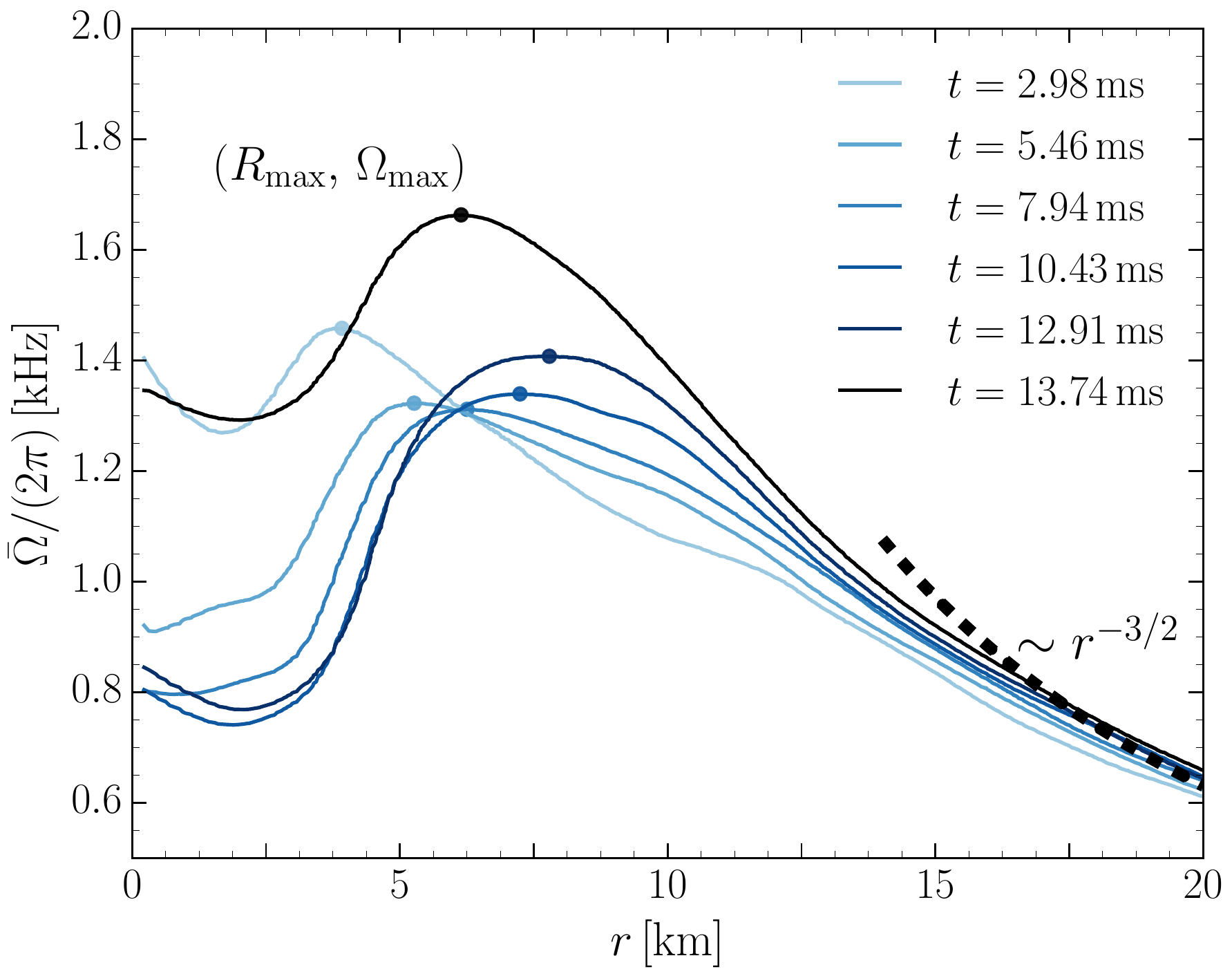}
  \caption{Averaged fluid angular velocity $\bar{\Omega}(r)/(2\pi)\,{\rm
      kHz}$ on the equatorial plane for the \texttt{ALF2-M135} binary as
    averaged at different times and with intervals of length $\Delta t
    =1\ms$. Shown as a thick dashed black line is a reference profile
    {scaling like $r^{-3/2}$}.}\label{fig:Omega-dt1-ALF2-1350}
\end{center}
\end{figure}

%-------------------------------------------------------
\subsection{Angular-velocity evolution: azimuthal averages}
\label{sec:omegaav}
%-------------------------------------------------------

As discussed before, for high-mass binaries  the angular-velocity
distribution exhibits an $m=2$ deformation that persists over long
timescales (this is however not the case
for low-mass binaries; see Sec. \ref{sec:lmb}). Because the
spacetime reaches rather quickly a stationary evolution and the
deformation is progressively washed out, it is reasonable to consider
azimuthal averages that, by reducing the problem to a one-dimensional
one, can help compare angular-velocity profiles across different
EOSs. Furthermore, as we will discuss in Sec. \ref{sec:lmb},
the approximation of an azimuthal average becomes increasingly good as
the mass of the system decreases and the $m=2$ deformation is more
rapidly lost. 

We define therefore the time- and azimuthally averaged angular velocity
$\bar{\Omega}(r,t)$ as
\begin{equation}
\bar{\Omega}(r,t) := \int_{t-\Delta t/2}^{t+\Delta t/2} \int_{-\pi}^\pi
\Omega(r,\phi,t') \, d\phi \, dt ' \,. \label{eq:Omegataverage}
\end{equation}
and show its evolution in Fig. \ref{fig:Omega-dt1-ALF2-1350} for the
\texttt{ALF2-M135} binary. Note that to better illustrate the time
dependence, a small time averaging domain ($\Delta t =1\ms$) has
been used. Six representative time segments, which span almost the whole
HMNS lifetime from merger to gravitational collapse, are
visualized in Fig.~\ref{fig:Omega-dt1-ALF2-1350}.

Soon after merger (\ie for $t \lesssim 3\ms$), the angular-velocity
profile varies considerably as angular momentum is transferred from the
central region to the outer layers (light-shaded blue lines). As a
result, the angular velocity decreases significantly in the inner
regions, quickly creating a large gradient with the more rapidly rotating
outer layers. The time variation of the azimuthal average is much smaller
as time progresses and the HMNS reaches a stationary configuration. Blue
curves with increasing shading in Fig.~\ref{fig:Omega-dt1-ALF2-1350} show
the rotation profiles for later time segments; clearly the qualitative
global structure of all of these curves is very similar. While the inner
parts ($r \lesssim 3\km$) of the HMNS rotate rather slowly ($\sim 1.0 $
kHz), a sharp increase takes place in a narrow region between $4\km$ and
$5.5\km$, resulting in an absolute maximum value at radii between 7-8
km. For larger radii, the angular velocity decreases monotonically
tending to a {$r^{-3/2}$} law for $r \gtrsim 15\km$ (see discussion
in Sec. \ref{sec:qub}).

The black curve in Fig.~\ref{fig:Omega-dt1-ALF2-1350} describes the
rotation profile of the HMNS at the brink of collapsing to a Kerr black
hole ($t \simeq 13.7\ms$) and can therefore be taken as the stationary
azimuthally averaged angular-velocity profile. About two milliseconds
before black-hole formation, the angular-velocity profile increases and
the position of the maximum moves inwards as a result of angular-momentum
conservation.

%-------------------------------------------------------
\subsection{Temperature evolution}
%-------------------------------------------------------

Reference \cite{Kastaun2016} has recently studied in great detail the
temperature distribution and the fluid trajectories of a HMNS produced by
a binary neutron star merger obeying the Shen, Horowitz and Teige EOS
\cite{ShenG2010,ShenG2011}. Although Ref. \cite{Kastaun2016} concentrates
on one EOS only, their results are in very good agreement with those
presented so far.  We next make a closer comparison with
\cite{Kastaun2016} by considering the evolution of a quantity we have not
yet discussed, namely, the fluid temperature. In particular, we will
concentrate on the binary \texttt{LS220-M132}, which is the only one of our
set described by an EOS with a consistent temperature, where the
simulations have been performed with the \texttt{WhiskyTHC} code
\cite{Radice2013b,Radice2013c}.

\begin{figure}%[b]
\begin{center}
\includegraphics[width=\columnwidth]{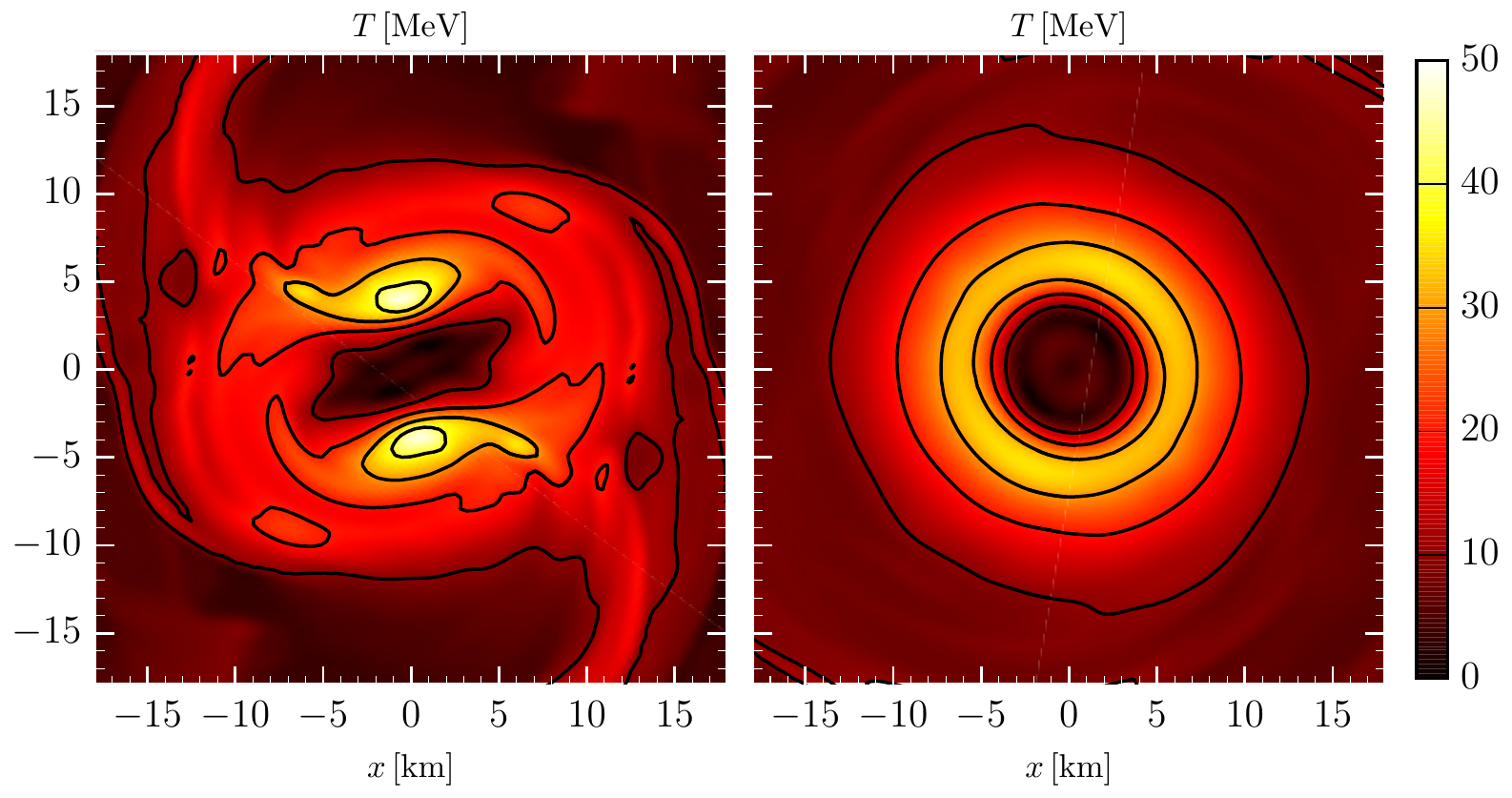}
\caption{Distributions on the $(x,y)$ plane and in a corotating frame of
  the temperature for the \texttt{LS220-M132} binary at $t=6.71\ms$ (left
  panel) and at $t=23.83\ms$ (right panel). The isocontours correspond to
$T= \{10,20,30,40,50\}$ MeV. Note the presence in the left
  panel of two hot spots, which do not coincide with the maximum
  rest-mass density (see also Fig.~\ref{fig:tracers}, which reports other
  quantities relative to this binary).}\label{fig:Temp-LS220}
\end{center}
\end{figure}

Figure~\ref{fig:Temp-LS220} shows the equatorial distributions of the
temperature at $t=6.71\ms$ (left panel) and at $t=23.83\ms$ (right panel;
note that the representative \texttt{ALF2-M135} binary has already
collapsed at this time) when observed in a ``corotating frame'', that is,
in a frame that is rotating at a frequency that is half of the
instantaneous gravitational-wave frequency. In agreement with
Ref. \cite{Kastaun2016}, two ``hot spots'' are found after the early
post-merger phase (see left panel of Fig.~\ref{fig:Temp-LS220}) which
remain stable for approximately 12 ms. The fluid trajectories (not
reported in Fig.~\ref{fig:Temp-LS220} but analysed in
Sec.~\ref{sec:tracers}) indicate that the hot spots also represent
vortices around which fluid elements rotate. Furthermore, the
temperature distribution on the equatorial plane bears a remarkable
similarity with the corresponding distribution of the angular velocity as
reported in Fig. \ref{fig:ALF2-M135-rho}. In particular, the position of
the hot spots overlaps closely with the position of the maxima in the
angular-velocity distribution. This is not surprising as in these regions
the fluid flow has the largest shear and compression, which are
ultimately responsible for the local increase of the temperature. 

\begin{figure*}
\begin{center}
\includegraphics[width=2\columnwidth]{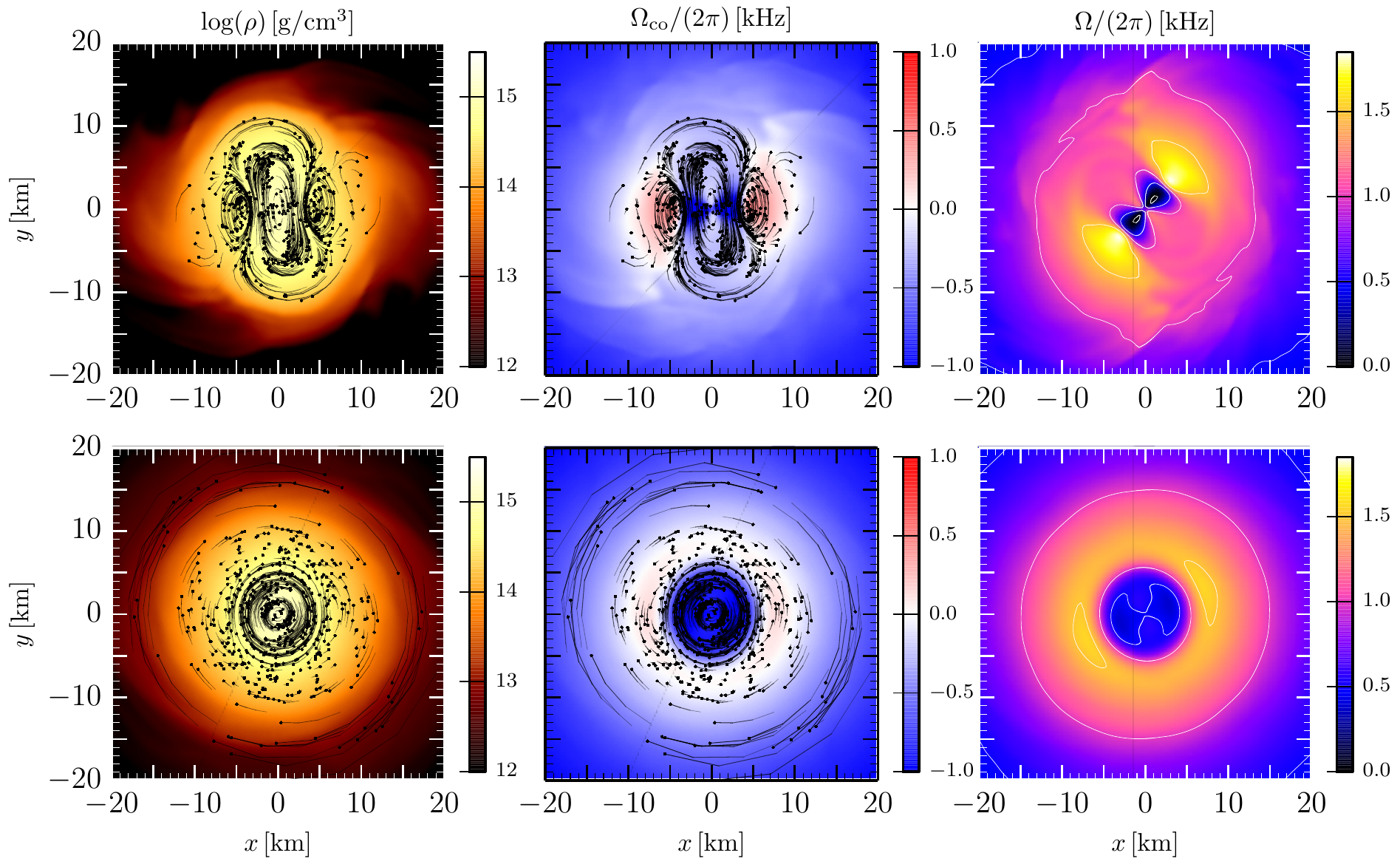}
\caption{Distributions on the $(x,y)$ plane and in the corotating frame
  of the rest-mass density (left panels), and of the angular velocity
  (middle panels) for the \texttt{LS220-M132} binary, where $\Omega_{\rm
    co} := \Omega - \Omega_{\rm f}$ where \lrn{$\Omega_{\rm f} = 1.368 $
    kHz (top), $1.425\,{\rm kHz}$ (bottom).}  Shown instead in the right
  panels is the distribution of the angular velocity, $\Omega$, in the
  Eulerian frame. The top row refers to $t=6.71\ms$, while the bottom one
  to $t=23.83\ms$. Also shown are portions of the flowlines of several
  tracer particles that remain close to the $(x,y)$ plane and for which
  we show only the final part of the flowlines (\ie for the last $\simeq
  0.285\ms$), using small dots to indicate the particle position at the
  time indicated in the frame. In addition, the initial parts of the
  trajectories have increasing transparency so as to highlight the final
  part of the trajectories.}
\label{fig:tracers}
\end{center}
\end{figure*}

Stated differently, regions of relatively smaller pressure (and rest-mass
density) should also coincide with regions of larger temperature, which
is what can be verified by comparing the rest-mass distribution in the
left panel of Fig.~\ref{fig:tracers}, with the temperature in the left
panel of Fig.~\ref{fig:Temp-LS220}, which refers to the same time. To the
best of our knowledge, this is the first time this explanation is
provided for the presence of the two ``hot spots''.

In addition to the hot spots, the temperature distribution shows local
increases along the edges of the $m=2$ density perturbation (again where
the fluid shear is largest in the corotating frame) and along the spiral
arms, where outward moving material is ejected dynamically from the
HMNS. Finally, we note that as time progresses and the HMNS reaches a
stationary state, the two hot spots disappear and the temperature
distribution reaches an axisymmetric pattern (see right panel of
Fig. \ref{fig:Temp-LS220}). Interestingly, however, the high-temperature
region is not the central one, which is slowly rotating and comparatively
colder, but, rather, an annular region at about 7-8 km from the center,
where the (axisymmetric) angular-velocity distribution varies more
rapidly. 

Although the discussion in this section has mainly focussed on the
\texttt{ALF2-M135} and \texttt{LS220-M132} binaries, the results are
qualitatively similar for all of the high-mass binaries of our sample. Of
course, small quantitative differences in terms of the maximum angular
velocity, or the precise position of the maxima of the $m=2$ deformations
in $\rho$ and $\Omega$ will depend on the stiffness of the EOS, but these
still obey the overall behaviour discussed so far; a more detailed
comparison of the angular-velocity profiles across the various EOSs is
presented in Sec. \ref{sec:qub}.

%%%%%%%%%%%%%%%%%%%%%%%%%%%%%%%%%%%%%%%%%%%%%%%%%%%%%%%%%%%%%%%%%%
\section{Low-mass binaries}
\label{sec:lmb}
%%%%%%%%%%%%%%%%%%%%%%%%%%%%%%%%%%%%%%%%%%%%%%%%%%%%%%%%%%%%%%%%%%

\begin{figure}%[t]
\begin{center}
\includegraphics[width=\columnwidth]{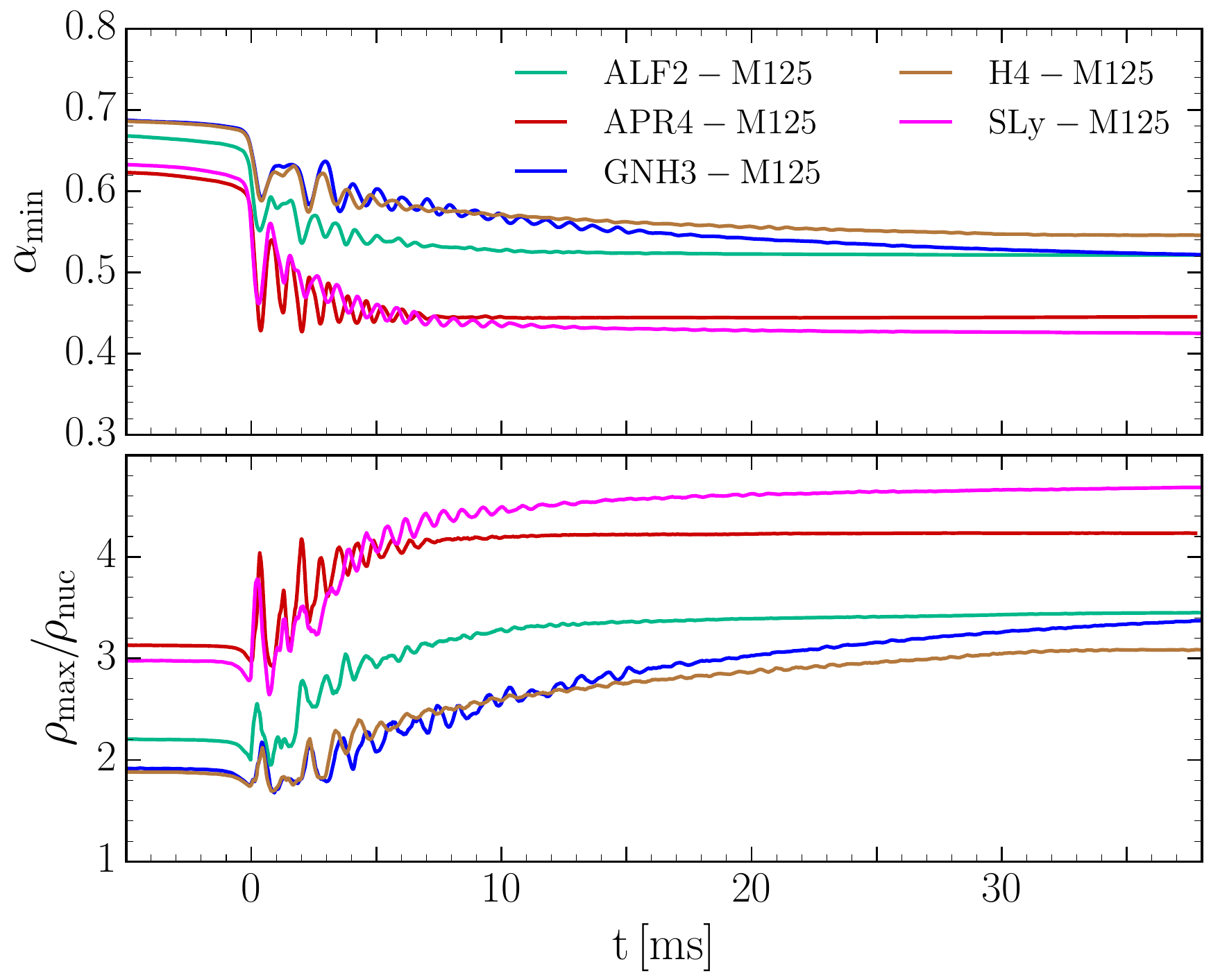}
\caption{Minimum value of the lapse function $\alpha_{{\rm {min}}}$
  (upper panel) and maximum of the rest-mass density $\rho_{{\rm {max}}}$
  in units of $\rho_{\rm nuc}$ (lower panel) versus time in milliseconds
  after the merger for the low-mass simulations; this figure should be
  contrasted with Fig.~\ref{fig:alphatimeH}.}\label{fig:alphatimeL}
\end{center}
\end{figure}

We turn now our attention to low-mass binaries. For this case we take
the ALF2 EOS as a representative EOS and concentrate on an equal-mass
binary with total mass $2\times 1.25\, M_\odot$, \ie
\texttt{ALF2-M125}. In analogy with Fig. \ref{fig:alphatimeH}, we show in
Fig.~\ref{fig:alphatimeL} the evolution of the minimum value of the lapse
function and of the maximum rest-mass density for our five EOSs. None of
the resulting HMNS collapses to a black hole within the simulated time
range because of the smaller initial mass of the binaries (they will
likely collapse on timescales of $\sim$100 ms; see \cite{Baiotti08} and
the discussion in Appendix of Ref. \cite{Rezzolla:2010}). It is also
interesting to note that the post-merger oscillations in the maximum
rest-mass density are all suppressed within about 10-15 ms from the
merger, with stiffer EOSs (\eg GNH3) requiring more time than the softer
ones (\eg APR4).

\begin{figure}%[b]
\begin{center}
\includegraphics[width=\columnwidth]{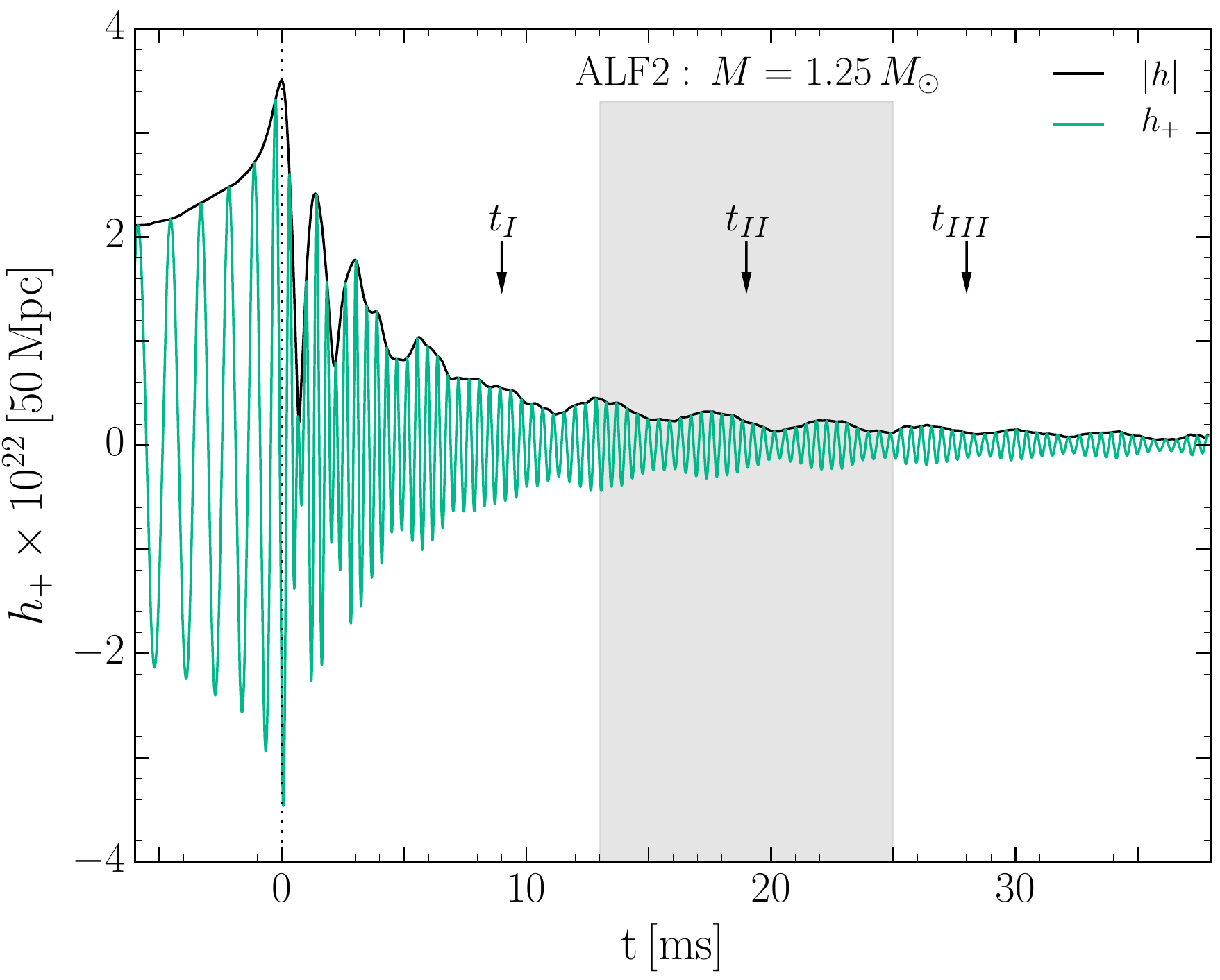} 
\caption{Gravitational-wave amplitude $|h|$ (black line) and strain
  amplitude in the $+$ polarisation $h_+$ (green line) for the
  \texttt{ALF2-M125} binary at a distance of 50 Mpc. Shaded in gray is
  the portion where a time average is performed, while the arrows
  indicate the times when representative distributions of the rest-mass
  density and angular velocity are shown in
  Fig.~\ref{fig:ALF2-M125-rho}. \lrn{Finally, the dotted vertical line
    marks the time of merger.} This figure should be contrasted with the
  equivalent one, Fig.~\ref{fig:GW-ALF2-M135}, for the high-mass
  binaries.}
\label{fig:GW-ALF2-M125}
\end{center}
\end{figure}

\begin{figure*}
\begin{center}
\includegraphics[width=2\columnwidth]{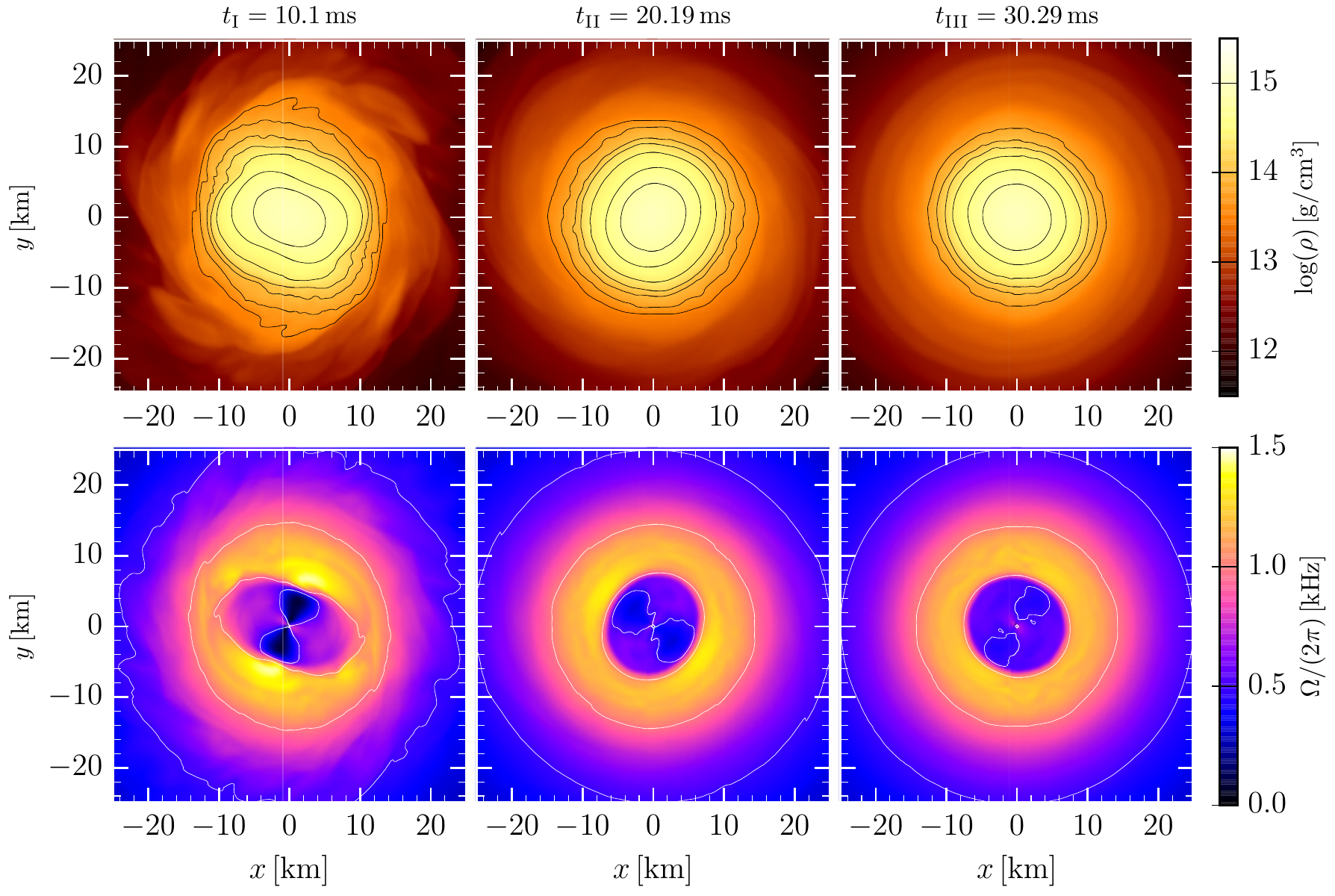}
\caption{Distributions of the rest-mass density (upper row, log scale)
  and of the fluid angular velocity (lower row) in the $(x,y)$ plane for
  the \texttt{ALF2-M125} binary at three different post-merger times as
  indicated in Fig.~\ref{fig:GW-ALF2-M125}. The isocontours
  have been drawn at $\log(\rho) = 13.6 + 0.2n$ (upper row) and $\Omega
  = \{0,0.5,1.0,1.5,2.0\} \, \rm kHz$ (lower row), $n \in
  \mathbb{N}$.}\label{fig:ALF2-M125-rho}
\end{center}
\end{figure*}

Similarly, we display in Fig.~\ref{fig:GW-ALF2-M125} the emitted
gravitational waves of the late inspiral, merger and post-merger, which
should be contrasted with the corresponding emission for the high-mass
binary shown in Fig.~\ref{fig:GW-ALF2-M135}. Note that the
gravitational-wave amplitude peak at the merger is comparable with that
of the high-mass run, but also that within the first 10 ms after merger
it decreases considerably to become only about 20\% of that in the late
inspiral. This is due to the rapid disappearance of the non-axisymmetric
deformation of the rest-mass density in the HMNS, which attains an almost
axisymmetric distribution within $\simeq 20\ms$ after the merger (see
upper panels in Fig.~ \ref{fig:ALF2-M125-rho}). The three different times
indicated in Fig. \ref{fig:GW-ALF2-M125} refer to $t=1/4$, $1/2$ and
$t=3t_{\rm fin}/4$, where $t_{\rm fin} = 40.38\ms$ is the time when the
simulation is terminated.

Figure \ref{fig:ALF2-M125-rho} reports the distributions on the
equatorial plane of the rest-mass density (upper panels) and of the
angular velocity (lower panels) at the three different times indicated in
Fig. \ref{fig:GW-ALF2-M125}. This figure should be compared with
Fig. \ref{fig:ALF2-M135-rho}, which refers to a high-mass binary of the
same (ALF2) EOS. Note that for $t\lesssim 16\ms$, the rotation profile
(see lower left panel in Fig.~\ref{fig:ALF2-M125-rho}) shows the same
qualitative structure as for the high-mass case, even though the overall
values of $\Omega$ are somewhat lower (note the different color scale in
Figs.~\ref{fig:ALF2-M135-rho} and \ref{fig:ALF2-M125-rho}). Although in a
weaker form, the low-mass binary also shows the 90-degree shift between
the $m=2$ deformation in the rest-mass density and in the angular
velocity, which we have discussed in Sec.~\ref{sec:omega2D} in terms of
the manifestation of the Bernoulli theorem [\cf
  Eq. \eqref{eq:bern0}]. Furthermore, for $t \gtrsim 16\ms$ (see middle
and right panels in Fig.~\ref{fig:ALF2-M125-rho}) the rest-mass density
and the rotation profile have reached a stationary state in which a small
$m=2$ perturbation is still present, but is subdominant when compared to
the overall axial symmetry. The inner part of the HMNS ($r \leq 6\km$) is
where the rest-mass density is the largest, but is also rotating rather
slowly ($\Omega \simeq 0.5\,{\rm kHz}$); this region is much broader than
in the high-mass binary and the sharp transition to a {$r^{-3/2}$}
outer profile takes place at a larger radius ($6.5 \lesssim r \lesssim
8.5$ km). The panels on the right column of Fig. \ref{fig:ALF2-M125-rho}
clearly indicate that at later times the HMNS has reached a high degree
of axial symmetry, although not a complete one, since gravitational waves
are still being emitted (\cf Fig. \ref{fig:GW-ALF2-M125}).

As discussed in Sec. \ref{sec:hmb} for the high-mass binaries, here too
we can comment that the analysis carried out on the equatorial plane
remains qualitatively valid also at nonzero elevations, with the
rest-mass density and rotation profiles not varying significantly for $z
\lesssim 9\km$, and maintaining the overall axisymmetry. Furthermore, as
the HMNS reaches rest-mass densities that are comparable with that of the
outgoing wind, for \lrn{$z \gtrsim 10$} km, the angular velocity
increases in the central regions, where it has the largest values.

Finally, we conclude this section remarking that much of the properties
discussed so far for the \texttt{ALF2-M125} binary remain qualitatively
true for all of the low-mass binaries considered in our sample of
models. Once again, small quantitative differences do appear when
considering the maximum angular velocity, or the precise position of the
maxima of the $m=2$ deformations in $\rho$ and $\Omega$, which obviously
depend on the stiffness of the EOS. However, the overall behaviour
discussed so far can be taken to be representative for binaries with
these masses and a large class of EOSs.

%%%%%%%%%%%%%%%%%%%%%%%%%%%%%%%%%%%%%%%%%%%%%%%%%%%%%%%%%%%%%%%%%%
\section{Tracer particles evolution}
\label{sec:tracers}
%%%%%%%%%%%%%%%%%%%%%%%%%%%%%%%%%%%%%%%%%%%%%%%%%%%%%%%%%%%%%%%%%%

To further strengthen the conclusions reached above about the physical
significance of the properties of the angular-velocity distribution
discussed in the previous section, we present below a complementary
analysis making use of the flowlines tracked by massless tracer particles
that are advected in the flow (this is not the first time tracer
particles are used in fully general-relativistic simulations and recent
related work can be found in Refs. \cite{Mewes2016, Kastaun2016}). More
specifically, we concentrate on the binary \texttt{LS220-M132}, as
evolved with the \texttt{WhiskyTHC} code \cite{Radice2013b,Radice2013c}.
Details on our implementation of tracers \lrn{can be found in
  \cite{Mewes2016}, while subtleties} of the information that can be
derived from them will be presented in a related work \cite{Bovard2016}.

The upper panels of Fig. \ref{fig:tracers} report, for the
\texttt{LS220-M132} binary and at time $t=6.7\ms$, the rest-mass density
$\rho$ (left panel), the angular velocity $\Omega$ in the corotating
frame (middle panel) and in the Eulerian frame (right panel). Similarly,
the bottom row of panels in Fig. \ref{fig:tracers} shows the same
quantities, but for a later time of $t=23.8\ms$. The times shown are
representative ones but the dynamics for this EOS is qualitatively very
similar to those presented in Secs. \ref{sec:hmb} and \ref{sec:lmb}. The
only difference is that although the binary \texttt{LS220-M132} belongs
to the high-mass class, its evolution does not lead to a collapse to a
black hole over the timescale during which the simulations have been
carried out, \ie $\sim 27.3\ms$ after merger.

Also shown in Fig.~\ref{fig:tracers} are the flowlines of several tracer
particles that remain close to the $(x,y)$ plane (\ie with small velocity
in the $z$ direction) and for which we introduce a novel visualization
technique. More specifically, we show only the final part of the
flowlines (\ie for the last $\simeq 0.285\ms$),
using small dots to indicate the particle position at the time indicated
in the frame. Furthermore, the initial parts of the trajectories have
increasing transparency so as to highlight the final part of the
trajectories. This approach has at least two advantages. First, it
provides a measure of the linear velocity (faster tracers leave longer
tracks); second, the presence of the filled dots and the increasing
transparency allow one to read-off the direction of motion.

The top panels in Fig.~\ref{fig:tracers} show the dynamics of the fluid
in the inner parts of the HMNS and highlight that two distinct regions
can be identified. The first region is in the core of the HMNS, where
there is an ellipsoidal structure orthogonal to the angular-velocity
distribution. Within this ellipsoid, fluid elements essentially move
clockwise along isobaric surfaces, with linear velocities that are rather
small in the inner regions. In addition to this ellipsoidal motion, the
tracers also show the presence of two small ``vortices'', \ie regions of
increased vorticity in this frame, which also coincide with the regions
of highest angular velocity, and which border the areas where velocity
drops almost to zero (in this frame). Note that it appears that the
tracers that are ``trapped'' in these vortices where they remain without
traversing the boundary to the central ellipsoid and where they have an
inverse sense of rotation (counter-clockwise). \lrn{Stated differently,
  the vorticity distribution in the corotating frame would show two
  islands of vorticity with different signs, referring to clock and
  counterclock wise rotation. Tracers in one region do not migrate to the
  other region.} This is mostly the result of using a corotating frame in
a flow that is differentially rotating. It is quite intuitive, in fact,
that if $\Omega$ is positive but not uniform, the transformation to a
corotating frame amounts to a net subtraction of a positive amount of
angular velocity, hence leading to areas of now negative angular
velocity. This becomes more apparent when considering the corresponding
picture in the Eulerian frame, as shown in the right panels of
Fig. \ref{fig:tracers}.

The bottom panels of Fig.~\ref{fig:tracers}, on the other hand, refer to
a much later stage of the HMNS evolution (\ie $t=23.83\ms$) and clearly
show that by this time the HMNS has attained an almost axisymmetric
structure, combined with a much smaller $m=2$ perturbation. As discussed
in the previous cases, also here the rotational profile of the HMNS
contains an inner area with $r\lesssim 5\km$ which is rotating slowly and
almost uniformly at $\Omega \approx 0.5\,{\rm kHz}$, followed by a sharp
increase at $5\lesssim r\lesssim 7\km$ reaching a maximum value $\Omega
\approx 1.5\,{\rm kHz}$ at $r\approx8\km$ and decreasing continuously for
$r \gtrsim 8\km$\footnote{Note that the angular-velocity distribution in
  the lower central panel of Fig. \ref{fig:tracers} refers to the
  corotating frame and that this frame is rotating at half the angular
  frequency of the emitted gravitational waves, $\Omega_{_{\rm
      GW}}$. Because the maximum of the angular velocity $\Omega_{\rm
    max}$ is of the order of $\Omega_{_{\rm GW}}/2$ (\cf left panel of
  Fig. \ref{fig:OmegaO2_3p}), \lrn{the ring structure in this panel is
    approximately at zero angular velocity.}}. Since this behaviour
follows the one described previously for \lrn{hybrid EOSs}, it suggests
that both the rest-mass density and the angular-velocity distributions
are preserved when using a fully three-dimensional EOS and radiative
losses are taken into account. In turn, this confirms that an analysis
carried out with piecewise polytropes and a thermal component does not
introduce a bias in the results.

\begin{figure}
\begin{center}
\includegraphics[width=\columnwidth]{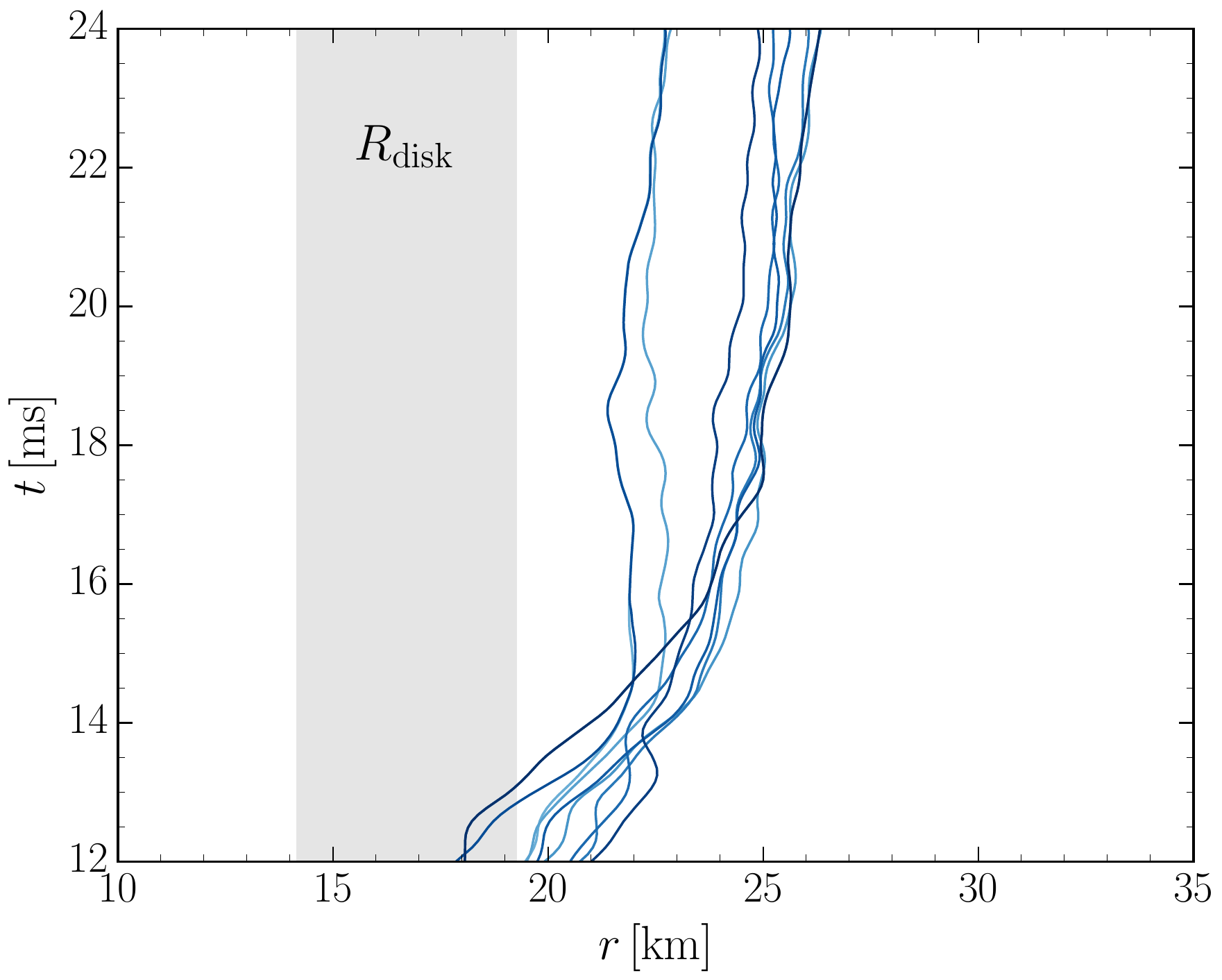}
\caption{Worldlines of selected tracers in the outer regions of the HMNS
  where the {angular frequencies scale like $r^{-3/2}$}. Note how in
  these regions the tracers remain at essentially constant radial
  coordinates; the gray-shaded area shows the region where
  {$\Omega(r) \propto r^{-3/2}$} (see Sec. \ref{sec:ap} for a
  definition and discussion).}\label{fig:circular_orbits}
\end{center}
\end{figure}

The tracers in the lower panels of Fig.~\ref{fig:tracers} further
illustrate the axisymmetric nature of the flow, with the fluid moving
along essentially circular orbits that are tangent to isobaric
surfaces. The quasi-circularity is shown in the spacetime diagram
reported in Fig. \ref{fig:circular_orbits}, where we show the worldlines
of selected tracers in the relevant region of the HMNS and after passing
them through a running-average window of $5\,{\rm ms}$ to remove the
high-frequency jitter. Note that after the transient period, where
angular momentum is transferred out and particles move to lower rest-mass
density regions, the tracers remain at essentially constant radial
coordinates.

It is interesting to note that because the
angular velocity refers to the corotating frame and because this frame is
rotating at half the gravitational-wave frequency ($\Omega_{_{\rm GW}}/2
\approx \Omega_{\rm 2}/2 \approx \Omega_{\rm max}$, see
Fig.~\ref{fig:OmegaO2_3p} of Section \ref{sec:qub}), tracers in the inner
and outer regions of the HMNS will be both rotating clockwise while
tracers belonging to the intermediate regions are almost at rest (white
regions). Only those tracers that are trapped in the vortices are moving
counter-clockwise (light red regions).

To recap, the analysis of the motion of tracer particles discussed in
this section indicates that the angular-velocity distributions presented
so far are not contaminated by gauge effects, but rather reflect
physically meaningful quantities. The angular-velocity $\Omega$ is
defined directly in terms of the lapse and shift, which are
gauge-dependent quantities and could potentially change the behaviour of
the angular-velocity depending on the gauge selected. The evolution of
the tracers are comoving with the fluid elements \cite{Bovard2016} and
thus they follow the fluid evolution. As shown, they exhibit behaviour as
predicted by the gauge-dependent quantities and illustrate the robustness
of our results to the choice of gauge. An additional use of the tracers,
explained in detail in Appendix \ref{sec:appendix_a}, will further
illustrate the origin of the phase offset demonstrated in
Fig.~\ref{fig:ALF2-M135-rho}.

%%%%%%%%%%%%%%%%%%%%%%%%%%%%%%%%%%%%%%%%%%%%%%%%%%%%%%%%%%%%%%%%%%
\section{``Quasi-universal'' behaviour}
\label{sec:qub}
%%%%%%%%%%%%%%%%%%%%%%%%%%%%%%%%%%%%%%%%%%%%%%%%%%%%%%%%%%%%%%%%%%

%-------------------------------------------------------
\subsection{Averaged profiles}
\label{sec:ap}
%-------------------------------------------------------

A particularly interesting result of our analysis emerges when comparing
the time- and azimuthally averaged profiles of the angular velocity
across the various EOSs and masses. We recall that the averaging
procedure has been discussed in Sec.~\ref{sec:omegaav} and requires a
proper choice of the integration interval, which is different in the case
of high-mass binaries (where the HMNS collapses to a black hole) and that
of low-mass binaries (where the HMNS survives through all the simulated
time). To maintain a certain consistency across these two cases, we have
performed the time averages across a time interval $\Delta t$ [see
  Eq.~\eqref{eq:Omegataverage}] centered around $t_{\rm fin}/2$ and with
extent $t_{\rm fin}/3$, where $t_{\rm fin}$ corresponds either to the
time of black-hole formation (in the case of high-mass binaries) or to
the final time of the simulation (in the case of low-mass
binaries). These time intervals have been indicated with gray-shaded
regions in Figs. \ref{fig:GW-ALF2-M135} and \ref{fig:GW-ALF2-M125},
respectively \lrn{(see Appendix \ref{sec:appendix_b} for a discussion on
  the impact of resolution on the lifetime of the HMNS and on the
  averages).}

\begin{figure}
\begin{center}
\includegraphics[width=\columnwidth]{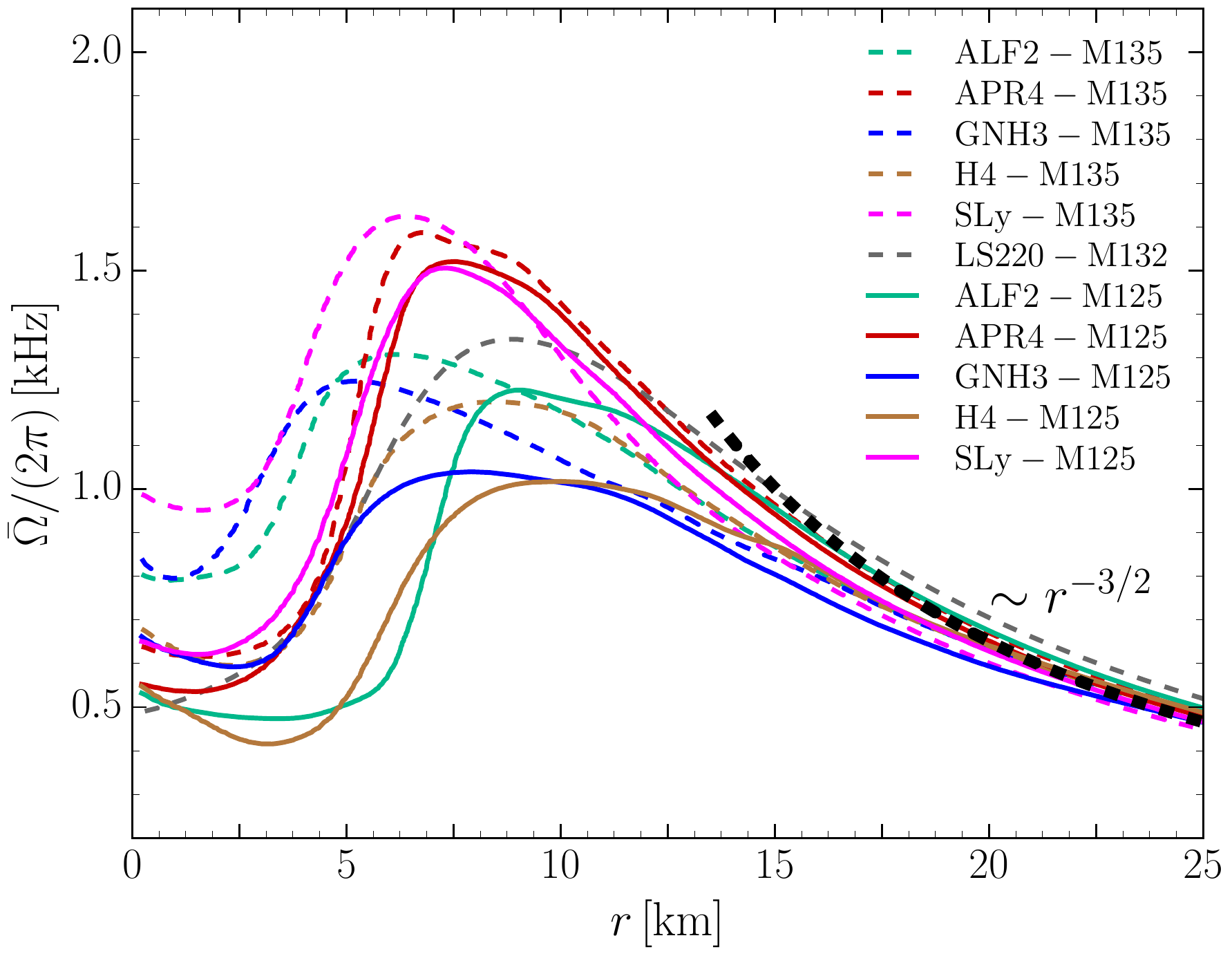}
\caption{Comparison of the time- and azimuthally averaged rotation
  profiles for different EOSs. Solid curves show the profiles for
  high-mass runs ($M=1.35 M_\odot$), whereas dashed curves refer to
  low-mass simulations ($M=1.25 M_\odot$). Shown as a thick dashed black
  line is a reference profile {scaling like
    $r^{-3/2}$}.}\label{fig:OmegaRadius}
\end{center}
\end{figure}

Figure~\ref{fig:OmegaRadius} shows the results of the time- and
azimuthally averaged angular-velocity profiles as described above and as
computed for the high and low-mass simulations with different
EOSs. Besides being in good agreement with the results of previous works
where a much smaller number of EOSs was investigated
\cite{Kastaun2014,Kastaun2016}, the overall behaviour of the
angular-velocity profiles shown in Fig.~\ref{fig:OmegaRadius} hints at a
behaviour that is only weakly dependent on the EOS and could therefore be
considered ``quasi-universal''.

\begin{figure}
\begin{center}
\includegraphics[width=\columnwidth]{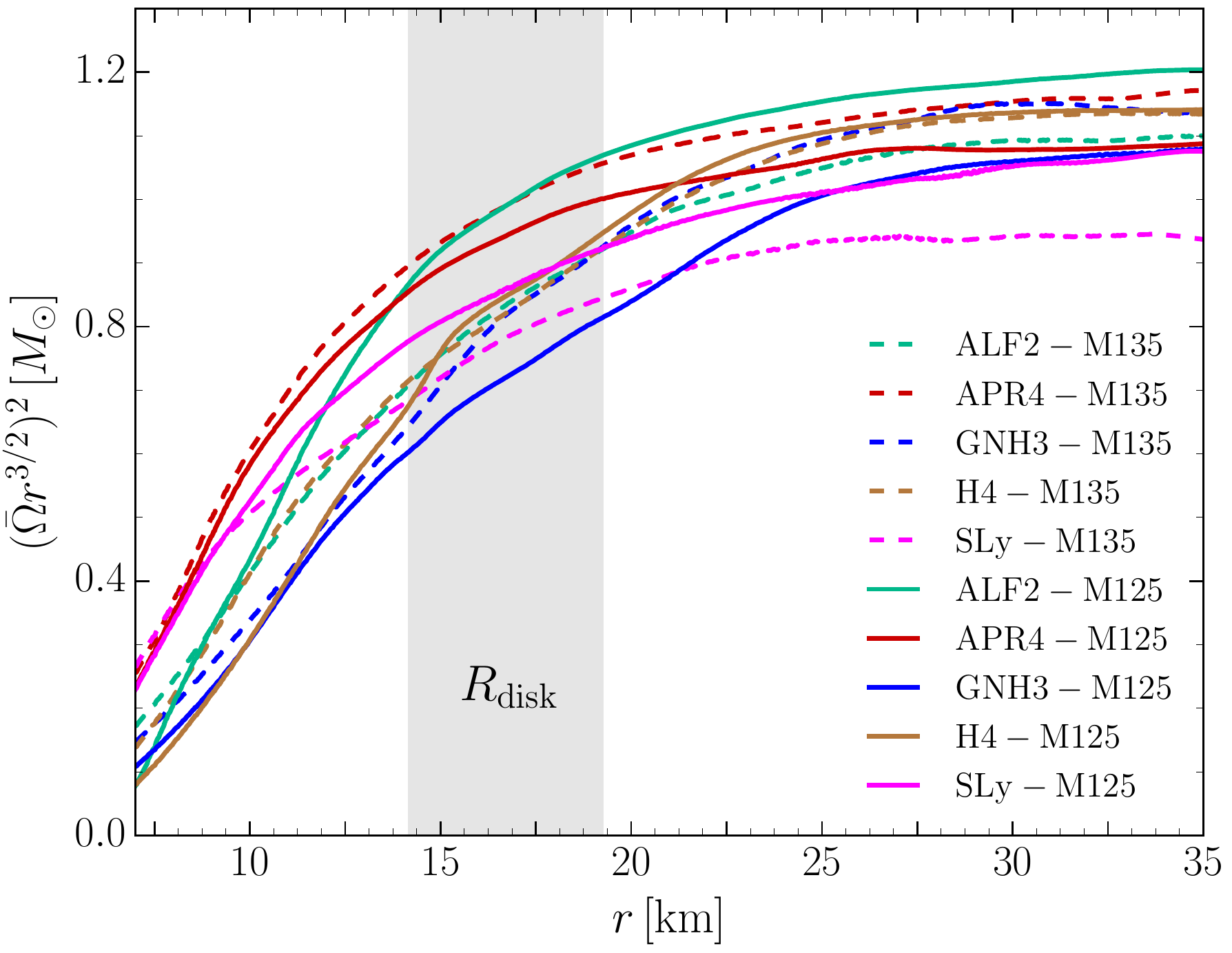}
\caption{Radial dependence of the quantity $\chi(r) := \bar{\Omega}^2
  r^3$ for all of the simulated binaries. Note that all profiles reach an
  almost constant value for $r\gtrsim 25\km$. The gray-shaded area shows
  the region where the flow starts to {having angular frequencies
    scaling like $r^{-3/2}$}; see Table \ref{tab:results} for the exact
  values of {$R_{\rm disk}$} for the various
  EOSs.}\label{fig:Kepler1}
\end{center}
\end{figure}

More specifically, all profiles show the presence of a slowly and
essentially uniformly rotating inner core, which is then joined by a
rapidly rotating outer region, hence decreasing outwards as $r^{-3/2}$,
and where the rest-mass density is that typical of neutron-star crusts
(see also Fig.~\ref{fig:Kepler1} and the discussion below). The
transition between the slow inner core and the rapidly rotating exterior
takes place across a narrow region which is only $3-4\km$ wide. It is
across this layer that shear forces are the largest and consequently
local temperature increases are present (see
Fig.~\ref{fig:Temp-LS220}). This differential rotation profile is rather
different from the one normally considered in the literature, \ie the
$j-{\rm constant}$ law that has been explored in the past both in
equilibrium configurations \cite{Shibata05c, Shibata06a, Ansorg2009,
  Galeazzi:2011, Studzinska2016, Gondek2016} and in dynamical ones
\cite{Baiotti06b, Manca07, Camarda:2009mk, Corvino:2010, Giacomazzo2011,
  Kiuchi2012b, Franci2013, Kaplan2013, Siegel2013, Siegel2014,
  Loeffler2015}.

Although very robust, small differences do appear within this
``quasi-universal'' behaviour. In particular, the spatial size of the
slow inner core depends both on the EOS and on the initial mass of the
binary, with smaller-mass binaries having in general larger slower cores;
as an example, for the high-mass run of the GNH3 and SLy EOS the slowly
rotating core is rather small ($r\lesssim 4\km$) while for the low-mass
run of the ALF2 EOS it extends up to $r \lesssim 7\km$.  In addition, the
maximum angular velocity attained in the 
{outer} region also depends weakly on the EOS, being naturally higher
for soft EOSs and smaller for stiff EOSs.

\begin{figure*}
\begin{center}
\includegraphics[width=2\columnwidth]{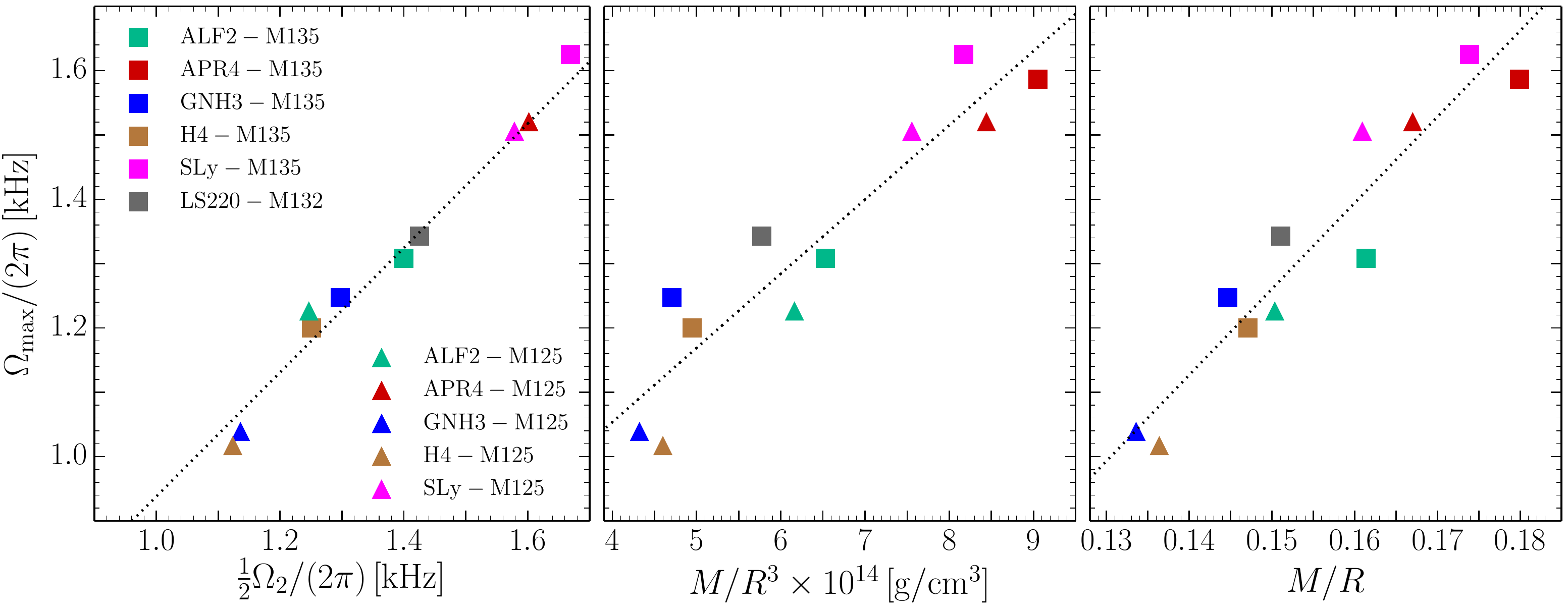}
\caption{Left panel: maximum value $\Omega_{\rm max}$ of the time- and
  azimuthally averaged rotation profiles (see Fig.~\ref{fig:OmegaRadius})
  as a function of (half of) the main gravitational-wave frequency of the
  emitted by the HMNS $\Omega_2$. Middle panel: $\Omega_{\rm max}$ as a
  function of the average rest-mass density $\bar{\rho}:=M/R^3$ relative
  to the initial stellar models. Right panel: $\Omega_{\rm max}$ as a
  function of the initial stellar compactness ${\cal C} :=
  M/R$.}\label{fig:OmegaO2_3p}
\end{center}
\end{figure*}

Inspired by the approach suggested in Refs. \cite{Bauswein2012a,
  Bauswein2012, Takami:2014, Takami2015, Rezzolla2016}, we next discuss
how to relate the ``quasi-universal'' features of the averaged
angular-velocity profiles with some of the properties of the merging
neutron stars, such as the mass and radius, when they are at infinite
separation.  Such correlations are summarised in
Fig.~\ref{fig:OmegaO2_3p}. We start by considering how the maximum
angular velocity $\Omega_{\rm max}$ of the averaged profiles relates to
the angular frequency corresponding to the largest peaks of the
post-merger power spectral density $\Omega_2 := 2\pi f_2$ (left panel of
Fig.~\ref{fig:OmegaO2_3p}; see \cite{Takami:2014, Takami2015,
  Rezzolla2016} for a definition and discussion of the various
frequencies of the post-merger signal). This frequency is customarily
interpreted as twice the spinning frequency of the $m=2$-deformed HMNS
\cite{Stergioulas2011b,Takami:2014} and is therefore not surprising that
it should tightly correlate with the maximum angular frequency of the
averaged angular-velocity profiles. What is less obvious is that the
value of $\Omega_{\rm max}$ reported in Fig.~\ref{fig:OmegaO2_3p} does
not change significantly with time and is therefore also the value of
$\Omega_{\rm max}$ at the end of the simulations, when the $m=2$
deformation has either been washed out or is small. Stated differently,
when the HMNS has reached an almost axisymmetric configuration, the
gravitational-wave frequency $\Omega_{_{\rm GW}} \simeq \Omega_2$ can
still be used to measure the maximum angular velocity of the fluid. In
addition, it is worth remarking that the correlation shown between
$\Omega_{\rm max}$ and $\Omega_2$ provides an additional confirmation
that the measurements of the angular-velocity distributions discussed in
the previous section (see discussion in Sec. \ref{sec:hmb}) is physically
meaningful. While in fact $\Omega_{\rm max}$ is gauge dependent,
$\Omega_2$ is one of the few gauge-independent quantities of our
simulations.

Shown instead in the central and right panels of
Fig. \ref{fig:OmegaO2_3p} are the correlations of $\Omega_{\rm max}$ with
the average rest-mass density $\bar{\rho} := M/R^3$ and the stellar
compactness ${\cal C} := M/R$. Also in this case, an essentially linear
correlation exists, which is easy to explain. Stiffer EOSs will lead to
HMNSs that have comparatively larger radii and hence smaller average
densities; in turn, smaller angular velocities will be necessary to
attain a quasi-stationary hydrostatic equilibrium for the
HMNS. Furthermore, for a given EOS, low-mass binaries will have
comparatively smaller average densities, thus explaining why, for the
same EOS, binaries reported with triangles have systematically lower
averaged maximum angular velocities (\cf central panel of
Fig.~\ref{fig:OmegaO2_3p}). Finally, the line of arguments described
above for the average densities applies unmodified also for the stellar
compactness (\cf right panel of Fig.~\ref{fig:OmegaO2_3p}).

We have already commented, when discussing Fig.~\ref{fig:OmegaRadius},
that the fluid flow in the outer regions of the HMNS, \ie for $r \gtrsim
15\km$, exhibits a  profile scaling like
$r^{-3/2}$. To substantiate this claim we recall that if, at the time the
HMNS has reached a quasi-stationary configuration, we \lrn{assume} the
spacetime at sufficiently large radii to be \lrn{sufficiently close to
  that of a Kerr black hole, so that} the geodesic motion of fluid
elements on the equatorial plane will have orbital angular frequencies
given by Kepler's expression \cite{Rezzolla_book:2013}
\begin{equation}
\Omega_{{\rm {Kep}}}(r) = \frac{\sqrt{M}}{\sqrt{r^3} + a\sqrt{M}}\,,
\label{eq:Keppler}
\end{equation}
where $a:=J/M$ is the spin parameter of the Kerr black hole, $J$ is the
total angular momentum, and $M$ is the total gravitational mass. Of
course this approximation is very crude since the HMNS's spacetime is not
that of a Kerr black hole, indeed the low-mass models do not collapse,
and the fluid motion is not entirely geodetic, but it is also true that
this approximation is very helpful to characterise the fluid motions in
the outer regions. We can therefore introduce the quantity
\begin{align}
\chi(r) := \Omega^2 (r) r^3 \,,
\label{eq:chi} 
\end{align}
which would tend asymptotically to the mass of the black hole if $\Omega
= \Omega_{{\rm {Kep}}}$, if the motion was a geodetic one, and if this
was a Kerr spacetime; obviously, none of these conditions are actually
met here. However, as we discuss below, we can use the radial dependence
of this quantity to measure where the flow starts {having angular
  frequencies scaling like $r^{-3/2}$}.

This is shown in Fig. \ref{fig:Kepler1}, which reports the behaviour of
$\chi(r)$ for the binaries simulated. Clearly, all of the profiles
converge to a rather constant value for large radii, thus indicating that
indeed the low rest-mass density regions of the HMNS exhibit an {a
  flow with angular frequencies scaling like $r^{-3/2}$}. In order to
determine where {this happens,} we follow a very phenomenological
approach and compute the scale height of $\chi$, that is, $\chi'/\chi :=
(d\chi/dr)/\chi$. Since $\chi \to {\rm const.}$ for a {flow with
  $\Omega(r) \propto r^{-3/2}$}, $\chi'/\chi$ will be zero when this
happens. Of course, {$\chi \to {\rm const.}$} only asymptotically and
so we approximate the location of the {transition to a disk} as the
place where the monotonically decreasing function $\chi'/\chi$ reaches a
sufficiently small value, which we here choose to be $\chi'/\chi \leq
0.05$.

The importance of having a  {disk} in the outer
regions of the HMNS is at least twofold. First, since a
 flow {with $\Omega \propto r^{-3/2}$}
satisfies the Rayleigh criterion of rotating fluids against axisymmetric
perturbations\footnote{We recall that this classical-physics criterion
  can also be seen as requiring that the specific angular momentum
  $j=\Omega(r) r^2$ increases outwards for a stably rotating fluid
  configuration.}, the differentially rotating ``disk'' surrounding the
HMNS \lrn{will probably accrete onto the uniformly rotating core of the
  HMNS only on a dissipative timescale. Therefore, it is possible that it
  will not affect the long-term stability required} in the proto-magnetar
model for short gamma-ray bursts~\citep{Zhang2001, Metzger2008,
  Bucciantini2012} and the subsequent extended X-ray emission
\cite{Rezzolla2014b}. Second, once the core of the HMNS eventually
collapses to a rotating black hole, the presence of a certain amount of
mass on stable orbits will guarantee that the black hole will not be
``naked'', as suggested in Ref. \cite{Margalit2015}, but rather
surrounded by a torus, which would then lead to the potential formation
of a relativistic jet. We will discuss this point in more detail in the
following section.

%-------------------------------------------------------
\subsection{Mass in the ``disk''}
%-------------------------------------------------------

The distribution of rest mass in the HMNS is of great astrophysical
importance as it regulates the amount of mass that is ejected in the
merger and that can subsequently feed r-process nucleosynthesis and an
electromagnetic counterpart to the merger via the radioactive decay of
by-products of the r-process (\ie via a macronova) (see, \eg
\cite{Li:1998, Kulkarni2005, Metzger:2010, Roberts2011, Kasen2013,
  Barnes2013, Tanaka2013, Rosswog2014a, Grossman2014, Lippuner2015,
  Radice2016}). In addition, and as mentioned in the previous section,
the knowledge of the rest-mass distribution in the HMNS, and in
particular of the portion of it in {the disk}, is important to
determine how much of the HMNS will ``survive'' the process of
gravitational collapse of the HMNS to a black hole and end up in building
a torus around the black hole.

\begin{figure}%[b]
\begin{center}
\includegraphics[width=\columnwidth]{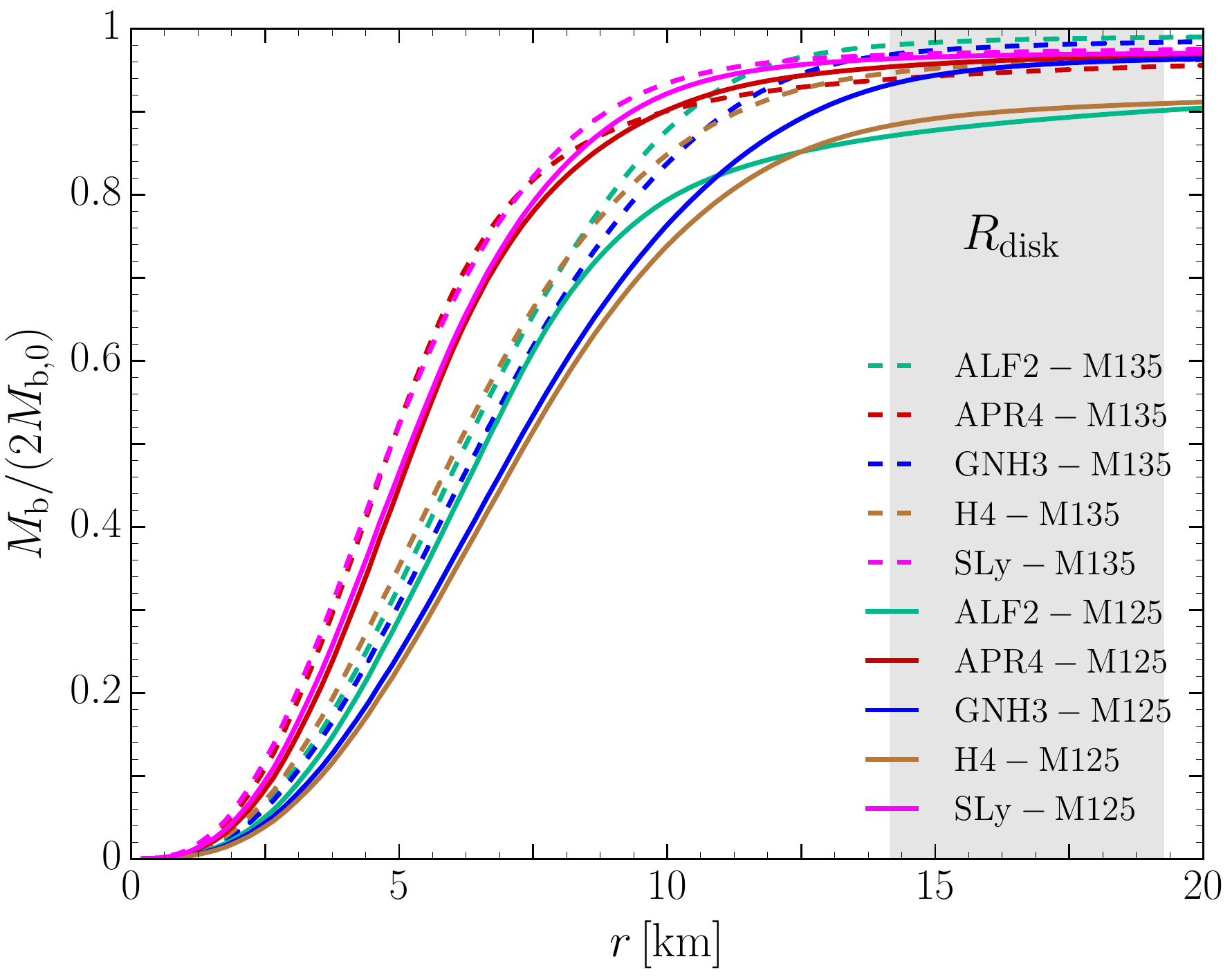}
\caption{Integrated rest mass $M_{{\rm b}}(r)$ as function of the radial
  coordinate and normalised with the total initial rest mass. The gray
  shaded area shows the region where the {disk starts}; see Table
  \ref{tab:results} for the exact values of {$R_{\rm disk}$} for the
  various EOSs.} \label{fig:Masses}
\end{center}
\end{figure}

As a result, we have computed the rest-mass distribution as a function of
the radial distance from the origin as
\begin{equation}
M_{{\rm b}}(\tilde{r}) := \int_0^{2\pi} \int_0^\pi \int_0^{\tilde{r}}
\sqrt{\gamma} \, W 
\rho \, {r}^2 \sin(\theta) \, d{r}\, d\theta\, d\phi\,,
\label{eq:RestMass} 
\end{equation}
with $W = \alpha u^t$ being the Lorentz factor. Figure \ref{fig:Masses}
illustrates the radial dependence of the total rest mass in the HMNSs
when normalised with the total initial rest mass $2\,M_{{\rm b,0}}$.
Note that both for the high- and low-mass binaries the distributions
refer to times which are in the middle of the averaging intervals in
Figs. \ref{fig:GW-ALF2-M135} and \ref{fig:GW-ALF2-M125} (\ie at
$t=t_{{\rm {BH}}}/2$ for the high-mass binaries and at $t=t_{{\rm
    {fin}}}/2$ for the low-mass binaries).

Irrespective of the EOS and initial mass, all of the rest-mass
distributions indicate that the rest mass $M_{{\rm b}}(r)$ does not
change significantly for $r \gtrsim 15\km$, so that the missing
amount of rest mass is the one that has been ejected dynamically soon
after the merger; \lrn{note that largest majority of this ejected matter
  is gravitationally bound and only a very small fraction of it will be
  ejected \textit{and} unbound \cite{Wanajo2014,Radice2016}}. Figure
\ref{fig:Masses} also shows a somewhat expected result, namely, that
binaries with softer EOSs (\eg APR4 or SLy) have considerably more
compact rest-mass distributions, reaching 80\% of the total within only
$r \lesssim 7.5\km$, quite independently of the initial mass of
the binary. By contrast, binaries with stiffer EOSs (\eg GNH3 or H4) have
less compact distributions, reaching 80\% of the total only for $r
\lesssim 10\km$, and a bit less for low-mass binaries.

\begin{table*}
\begin{footnotesize}
\begin{tabular}{l|c|c|c|c|c|c|c|}
\hline
\hline
model 
& $R_{\Omega_{\rm max}}$  & $\Omega_{\rm max}$  & $M_{{\rm b}, \Omega_{\rm max}}$
& $M_{{\rm b}, \Omega_{\rm max}}/(2\,M_{{\rm b,0}})$  & {$R_{\rm disk}$} 
& $M_{{\rm b, disk}}$ 
& $M_{{\rm b, disk}}/(2\,M_{{\rm b,0}})$  \\
& $[\mathrm{km}]$  & $[\rm kHz]$   & $[\Msun]$    
& $[\%]$ & $[\mathrm{km}]$ & $[\Msun]$ & $[\%]$ \\
\hline
\texttt{GNH3-M125} &  $ 7.92 $ & $ 1.04 $ & $ 1.56 $ & $ 57.89 $ & $ 18.96 $ & $ 0.10 $ & $ 3.79$  \\
\texttt{GNH3-M135} &  $ 5.19 $ & $ 1.25 $ & $ 0.97 $ & $ 33.12 $ & $ 19.27 $ & $ 0.05 $ & $ 1.65$   \\
\hline
\texttt{H4-M125}   & $ 9.98 $ & $ 1.02 $ & $ 1.99 $ & $ 73.70 $ & $ 17.08 $ & $ 0.26 $ & $ 9.66  $ \\
\texttt{H4-M135}   & $ 8.36 $ & $ 1.20 $ & $ 2.18 $ & $ 74.20 $ & $ 17.07 $ & $ 0.12 $ & $ 4.10  $ \\
\hline
\texttt{ALF2-M125} & $ 9.04 $ & $ 1.23 $ & $ 2.04 $ & $ 74.49 $ & $ 15.71 $ & $ 0.32 $ & $ 11.73 $ \\  
\texttt{ALF2-M135} & $ 6.20 $ & $ 1.31 $ & $ 1.46 $ & $ 49.35 $ & $ 16.39 $ & $ 0.04 $ & $ 1.35  $  \\
\hline
\texttt{SLy-M125}  &  $ 7.31 $ & $ 1.51 $ & $ 2.13 $ & $ 77.44 $ & $ 14.15 $ & $ 0.10 $ & $ 3.63  $ \\
\texttt{SLy-M135}  &  $ 6.43 $ & $ 1.63 $ & $ 2.16 $ & $ 72.17 $ & $ 15.05 $ & $ 0.09 $ & $ 3.01  $ \\
\hline
\texttt{APR4-M125} &  $ 7.57 $ & $ 1.52 $ & $ 2.15 $ & $ 78.49 $ & $ 14.56 $ & $ 0.12 $ & $ 4.43  $ \\
\texttt{APR4-M135} &  $ 6.76 $ & $ 1.59 $ & $ 2.30 $ & $ 76.40 $ & $ 14.70 $ & $ 0.18 $ & $ 5.85  $ \\
\hline
\hline
\end{tabular}
\caption{Summary of the HMNS properties. The various columns denote the
  radial position $R_{\Omega_{\rm max}}$ of the maximum of the averaged
  angular-velocity profiles $\Omega_{\rm max}$, the total rest mass
  inside $R_{\Omega_{\rm max}}$, \ie $M_{{\rm b}, \Omega_{\rm max}} :=
  M_{{\rm b}} (R_{\Omega_{\rm max}})$, the radial position {$R_{\rm
      disk}$} where the {disk starts}, and the total rest mass
  outside {$R_{\rm disk}$}, \ie $M_{\rm b, disk} := 2\,M_{{\rm b,0}}
  - M_{\rm b} (R_{\rm disk})$.
\label{tab:results}}
\end{footnotesize}
\end{table*}

Another useful measure of the rest-mass distribution is that of the mass
in the ``disk'', which we define to be the rest mass confined in the
region of the HMNS in {the disk}. More specifically, if {$R_{\rm
    disk}$} is the radial location (on the equatorial plane) where the
{disk starts} (\ie where $\chi'/\chi \leq 0.05$), then we define the
rest mass in the disk as $M_{{\rm b, disk}}:= 2\,M_{{\rm b,0}} - M_{{\rm
b}}$ ({$R_{\rm disk}$}). Another interesting notion of mass is
that that is inside the maximum value of the averaged angular-velocity
profile.  After defining $R_{\Omega_{\rm max}}$ as the location where the
angular-velocity profile reaches its maximum, and recalling that \ie
$R_{\Omega_{\rm max}} <$ {$R_{\rm disk}$}, we can calculate the mass
inside $R_{\Omega_{\rm max}}$ using the same expression
\eqref{eq:RestMass}; we refer to this mass as to $M_{b,\Omega_{\rm
    max}}$. The masses in the disk surrounding the quasi-uniformly
rotating inner core of the HMNS and that outside the angular-velocity
maximum are summarised in Table \ref{tab:results} for all of the binaries
simulated. Overall, they indicate that the mass outside $R_{\Omega_{\rm
    max}}$ and the mass outside {$R_{\rm disk}$} are rather similar
(the former being slightly larger since since $R_{\Omega_{\rm max}}
\lesssim$ {$R_{\rm disk}$}) and can be quite large, being almost
$0.3\,M_{\odot}$ for some of the soft-EOS low-mass binaries and of the
order of $0.1\,M_{\odot}$ for the other binaries. \lrn{Long-term
  angular-momentum transport and neutrino radiation will change
  quantitatively these values, but we do not expect them to change the
  qualitative picture that low-mass soft-EOS binaries will have
  comparatively larger disks.}

%-------------------------------------------------------
\subsection{Influence of the thermal component}
\label{sec:itc}
%-------------------------------------------------------

The final section of this paper is dedicated to assessing the impact of
the thermal component of the EOS on the results presented so far, further
extending the discussion made in Sec.~\ref{sec:tracers}. We recall, in
fact, that with the exception of the LS220 EOS, all of our EOSs do not
have a nuclear-physics thermal component and that thermal effects are
accounted for via a hybrid EOS in which an ideal-fluid contribution is
added to the total pressure (see Sec.~\ref{sec:eos} for details). The
choice of the adiabatic index $\Gamma_\mathrm{th}$ is somewhat arbitrary
(the only mathematical constraint being that $1 \leq \Gamma_\mathrm{th}
\leq 2$, but see discussion in~\cite{Rezzolla_book:2013}). Since the
value of $\Gamma_\mathrm{th}$ regulates the amount of thermal pressure
produced after merger and hence, to some extent, the equilibrium
properties of the HMNS, it is worth investigating its effects on the
rest-mass density and angular-velocity distributions.

\begin{figure}%[b]
\begin{center}
\includegraphics[width=\columnwidth]{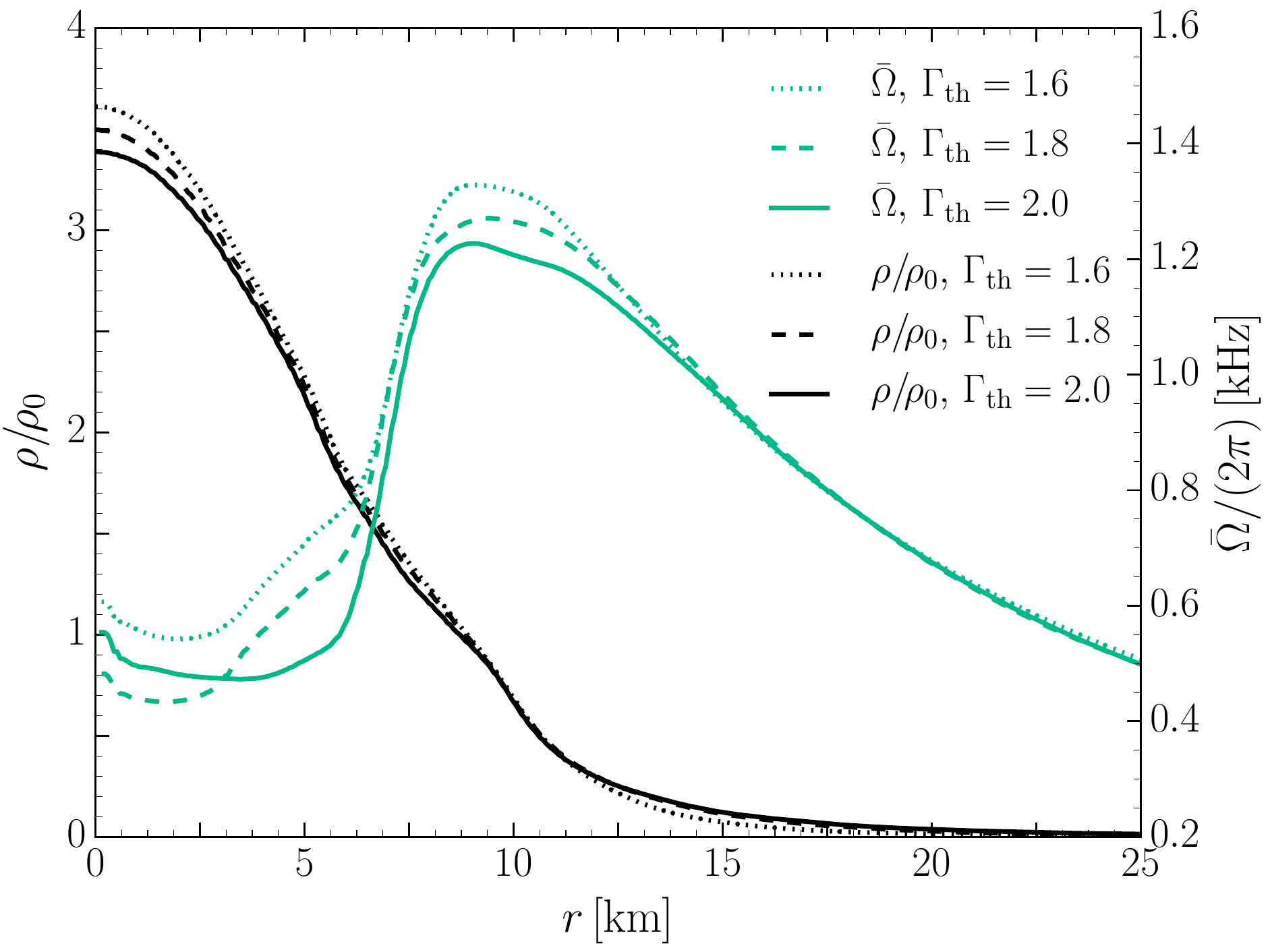}
\caption{Time- and azimuthally averaged rest-mass density profiles (black
  lines) and angular-velocity profiles (green lines) of the
  \texttt{ALF2-M125} binary for different values of the thermal adiabatic
  index $\Gamma_\mathrm{th}$. The time and azimuthal averages have been
  performed in the same manner as in
  Fig.~\ref{fig:OmegaRadius}.}\label{fig:OmegaRadius-gth1}
\end{center}
\end{figure}

This is shown in Fig.~\ref{fig:OmegaRadius-gth1} for the
\texttt{ALF2-M125} binary, but the results are similar for other
binaries. We indicate with green (black) solid, dashed and dotted lines
the averaged angular velocity (rest-mass density) profiles for
$\Gamma_\mathrm{th}=2.0, 1.8$ and $1.6$, respectively. The curves are
obtained after averaging in the azimuthal direction and over a time
interval [$1/3\,t_{\rm fin} , 2/3\,t_{\rm fin}$].  Figure
\ref{fig:OmegaRadius-gth1} shows that a larger value of
$\Gamma_\mathrm{th}$ yields a larger pressure support and hence prevents
the matter in the HMNS to reach large values of compression. This
explains why the maximum rest-mass density is larger for smaller values
of the thermal adiabatic index (see black lines). In turn, since a larger
pressure support implies that the HMNS is less compact (compressed) and
since the angular momentum is essentially the same for the binaries with
different $\Gamma_\mathrm{th}$ considered here, it is not surprising that
the maximum value of the averaged angular velocity increases as the
contribution of the thermal component is decreased (see green lines).

What is possibly more important to note is that the changes induced by
the different values of $\Gamma_\mathrm{th}$ are quantitative only and
also rather small, \ie with relative variations of $\lesssim 10\%$ in the
angular velocity. The qualitative behaviour, however, remains unchanged,
most notably, in the {disk}, thus removing the influence of the
thermal component of the EOS as a potential bias in our analysis.

%%%%%%%%%%%%%%%%%%%%%%%%%%%%%%%%%%%%%%%%%%%%%%%%%%%%%%%%%%%%%%%%%%
\section{Conclusions}\label{sec:sum}
%%%%%%%%%%%%%%%%%%%%%%%%%%%%%%%%%%%%%%%%%%%%%%%%%%%%%%%%%%%%%%%%%%

Establishing the long-term stability properties of astrophysical compact
objects produced in catastrophic events, such as in Type-II core collapse 
supernovae or in the merger of binary systems of neutron stars, is an old and
important problem. While it is clear that the large angular momentum that
these objects attain cannot be sustained via uniform rotation, far less
clear is what is the law of differential rotation that is reached in
quasi-stationary equilibria. More importantly, it is not yet known
whether this law depends sensitively on the EOS of the compact object or
is instead ``universal''.

Notwithstanding these conceptual obstacles, a large bulk of literature
has developed over the last decade to explore this problem in full
general relativity, either through the study of equilibrium
configurations or via numerical relativity simulations that produce these
objects dynamically. Works in the first class have commonly modelled
differential rotation through a particularly simple law expressing that
the specific angular momentum is constant on cylinders. In this $j-{\rm
  constant}$ law, angular velocity decreases \textit{monotonically} from
the center of the star and the degree of differential rotation is
expressed via a single dimensionless parameter. However, a number of
simulations of merging binary neutron stars have given evidence that the
angular-velocity profile of the HMNS produced at the merger is
characterized by a slowly rotating core and an envelope that rotates at
 frequencies {scaling like $r^{-3/2}$}. This is
very different from what is expected when using a $j-{\rm constant}$ law
of differential rotation.

To shed some light on these differences, and to obtain a comprehensive
picture of the rotational properties of HMNSs from binary neutron-star
mergers, we have carried out a large number of numerical simulations in
full general relativity of binary neutron stars described with various
EOSs and masses.  We have been able to confirm the earlier results of
Refs.~\cite{Shibata06a, Kastaun2014, Kastaun2016}, but, more importantly,
to show that the angular-velocity distribution shows only a modest
dependence on the EOS, thus exhibiting the traits of
``quasi-universality''. More specifically, the EOS-independent
angular-velocity distributions we find are characterized by an almost
uniformly rotating core and a  ``disk''. The rest
mass contained in such disk can be quite large, ranging from $\simeq
0.03\,M_{\odot}$ in the case of high-mass binaries with stiff EOSs, up to
$\simeq 0.2\,M_{\odot}$ for low-mass binaries with soft EOSs.

The presence of a {disk} in the outer regions of the HMNS implies
that the disk will only accrete onto the uniformly rotating core on a
dissipative timescale, thus not affecting the long-term stability of the
latter\footnote{\lrn{By long-term stability we here refer to a timescale
    which is much longer than the uncertainty in the lifetime of the HMNS
    due to a finite numerical resolution.}}. The final fate of this disk
when the lifetime of the HMNS is expected to be very large (\ie larger
than $10$ s and up to $10^4$ s \cite{Ravi2014}) is hard to assess through
self-consistent numerical simulations. However, its dynamics has been
recently conjectured within the ``two-winds'' model presented in
Ref. \cite{Rezzolla2014b} to explain the extended X-ray emission observed
in a class of short gamma-ray bursts. In essence, the expectation is that
because the material in the outer regions is on stable orbits, it could
be subject to a magnetorotational instability \citep{Velikhov1959,
  Chandrasekhar1960, Balbus1991} and hence behave as a standard accretion
disk onto a rapidly rotating magnetized star. In this case, the
differential rotation will not brake the rotation but transport angular
momentum outwards. A good fraction of the material in this disk will
therefore remain on {quasi-circular} orbits. Once the uniformly
rotating core has collapsed, the material in the disk that was not able
to accrete before because of the presence of the inner-core ``surface'',
will accrete onto the black hole on the timescale set by the most
efficient dissipative mechanism removing angular momentum.  Clearly, if
the mass in the torus is very small, then it will become difficult to
find the energy reservoir needed to launch and sustain the jet that is
required in the two-winds model. However, given that at least 10\% of a
solar mass is present in the disk after 20 ms after the merger, it is
sufficient that only 10\% of this mass is channelled into a torus around
the newly formed black hole to provide a sufficient amount of energy to
power a relativistic jet. More work is needed to fully explore the
consequences of this scenario and also to assess the impact that
neutrino-driven winds may have on the survival of the disk
\cite{Just2015}.

On a final note, we should also draw attention to various
  limitations of the work presented here. First, all of the binaries
  considered here have the same mass; although the masses in observed
  neutron-star binaries do not differ significantly, it is unlikely that
  they are exactly the same. Fortunately, it seems that this systematic
  bias in our sample may not be a serious one since the results presented
  in Ref. \cite{Ciolfi2017} for unequal-mass binaries show very similar
  angular-velocity profiles. Second, our simulations do not account for
  neutrino transfer; this is mostly because the neutrino-diffusion
  timescale is at least one order of magnitude larger than the one
  considered here \cite{Paschalidis2012} and a complete investigation of
  the rotational properties of the HMNS when varying the EOS and the
  neutrino transport are still prohibitive. Hence, we have preferred to
consider here a more controlled scenario in which we evaluate only the
impact of the EOS. Third, our simulations have neglected magnetic
  fields, even though we do expect magnetic fields to impact the dynamics
  of the HMNS by transferring angular momentum from the inner regions of
  the HMNS out to the more external ones. This transfer will take place
  on an Alfv\'en timescale, which is much longer than the one considered
  here. In this sense, the quasi-universal behaviour reported here should
  be taken as representative of the first few tens of milliseconds after
  the merger and recent work in Ref. \cite{Ciolfi2017} indicates that our
  results should remain valid also in the presence of magnetic fields and
  at least within $\sim 40\,{\rm ms}$ after merger.
The importance of magnetic fields in the post-merger dynamics has been
discussed in recent works~\cite{Rezzolla:2011, Kiuchi2014, Kiuchi2015a,
  Dionysopoulou2015, Palenzuela2015, Ruiz2016, Kawamura2016}.  During
merger a vortex sheet develops where the tangential component of the
velocity is discontinuous. Such a shear interface is unstable to
perturbations and can develop the so-called Kelvin-Helmholtz
instability. If the magnetic-field of the merging neutron stars is
poloidal, this instability may lead to an exponential growth of their
toroidal component~\cite{Price06, Baiotti08, Giacomazzo:2010,
  Rezzolla:2011, Kiuchi2014, Kiuchi2015a}, curling poloidal field lines
and generating a turbulent flow. Resolving the growth and saturation of
the Kelvin-Helmholtz instability is a challenging computational issue,
especially for realistic (pulsar-like) magnetic field strengths. The
latest, highest resolution simulations to date of~\cite{Kiuchi2015a} for
initial magnetic fields of moderate strength ($10^{13}$ G) show that the
amplification factor of the field is $\sim 10^3$ at $\sim 4\ms$ after
merger, with a saturation magnetic field-energy $\gtrsim 4\times 10^{50}$
erg, \ie~$\gtrsim 10^{-3}$ of the bulk kinetic energy of the merging
neutron stars. Although the magnetic energy is still much smaller than
the bulk kinetic energy and unable to produce significant changes in the
inspiral \cite{Giacomazzo:2009mp}, these results and the possibility that
even stronger amplification could be achieved, suggest the importance to
consider strong magnetic fields for modelling the post-merger evolution
of binary neutron stars. Finally, our investigation has been
  concentrated mostly on ``cold'' EOSs, where a thermal contribution has
  been added in a way that is ``ad-hoc'', albeit quite customary [\cf
    Eq. \eqref{EOS:full_a}]. It is therefore reassuring that two of the
  12 binaries simulated here, and for which a ``hot'' EOS has been
  employed, show a behaviour for the angular velocity that is very
  similar to that encountered for the cold EOSs. Overall, the caveats
  listed above represent strong motivations to further refine the study
  carried out here and advance our understanding of the rotational
  properties of the HMNSs produced in binary neutron-star mergers.

%%%%%%%%%%%%%%%%%%%%%%%%%%%%%%%%%%%%%%%%%%%%%%%%%%%%%%%%%%%%%%%%%%
\section*{Acknowledgements}
%%%%%%%%%%%%%%%%%%%%%%%%%%%%%%%%%%%%%%%%%%%%%%%%%%%%%%%%%%%%%%%%%%

We thank B. Mundim, N. G\"urlebeck, M. G. Alford, W. Kastaun 
%% and the anonymous reviewer
for useful discussions and comments. Support comes from the ERC Synergy
Grant ``BlackHoleCam'' (Grant 610058), from ``NewCompStar'', COST Action
MP1304, from the LOEWE-Program in HIC for FAIR, from the European Union's
Horizon 2020 Research and Innovation Programme (Grant 671698) (call
FETHPC-1-2014, project ExaHyPE), from JSPS KAKENHI grant (Grant
15H06813), from the Spanish MINECO (grants AYA2013-40979-P and
AYA2015-66899-C2-1-P), and from the Generalitat Valenciana
(PROMETEOII-2014-069). MH gratefully acknowledges support from the
Frankfurt Institute for Advanced Studies (FIAS) and the Goethe
University Frankfurt, while HS acknowledges the Judah M. Eisenberg
laureatus Professur endowment. The simulations were performed on SuperMUC
at LRZ-Munich, on LOEWE at CSC-Frankfurt and on Hazelhen at HLRS in
Stuttgart.

\appendix

%%%%%%%%%%%%%%%%%%%%%%%%%%%%%%%%%%%%%%%%%%%%%%%%%%%%%%%%%%%%%%%%%
\section{\lrn{On the Bernoulli constant}}
\label{sec:appendix_a}
%%%%%%%%%%%%%%%%%%%%%%%%%%%%%%%%%%%%%%%%%%%%%%%%%%%%%%%%%%%%%%%%%

As discussed in section \ref{sec:omega2D}, the evident $\pi/2$ phase
difference in the distribution of the angular velocity and of the
density, so that areas of low pressure (rest-mass density) are
accompanied by regions of large velocity, can be explained in terms of
the manifestation of the Bernoulli theorem. To show this, we will
consider a rather idealised description of the quasi-stationary
equilibrium of the HMNS and make a series of assumptions that will
simplify the mathematical treatment and hopefully improve our physical
understanding.

We therefore start by recalling that in relativistic hydrodynamics and
for a perfect fluid with four-velocity $\boldsymbol{u}$, the quantity
$h\,(\boldsymbol{u}\cdot\boldsymbol{\xi})$ is Lie-dragged along
$\boldsymbol{u}$ \cite{Rezzolla_book:2013}
\begin{equation}
\label{eq:bern0}
\mathscr{L}_{\boldsymbol{u}}
(h\,\boldsymbol{u}\cdot\boldsymbol{\xi})=0 \,,
\end{equation}
where $h:= (e+p)/\rho$ is the specific enthalpy, $e:=\rho(1+\epsilon)$ is
the total energy density, and $\boldsymbol{\xi}$ is a Killing vector of
the spacetime and also a generator of the symmetry obeyed by the fluid. A
direct consequence of Eq. \eqref{eq:bern0} is that in the case in which
the spacetime admits a timelike Killing vector, then the quantity
$\mathcal{B}:= h u_t$ is a constant of the fluid; this is the
general-relativistic extension of the classical Bernoulli theorem. 

Of course, the assumption of a stationary spacetime during the evolution
of the HMNS is not true and the HMNS is emitting gravitational waves,
which are the most evident manifestation of a non-stationary
spacetime. Yet, because at least energetically these modulations of the
spacetime are small when compared with the total bulk (kinetic) energy of
the system, we can consider the assumption not to be unreasonable even
though it is evidently not strictly true.

\begin{figure*}
\begin{center}
  \includegraphics[width=1.75\columnwidth]{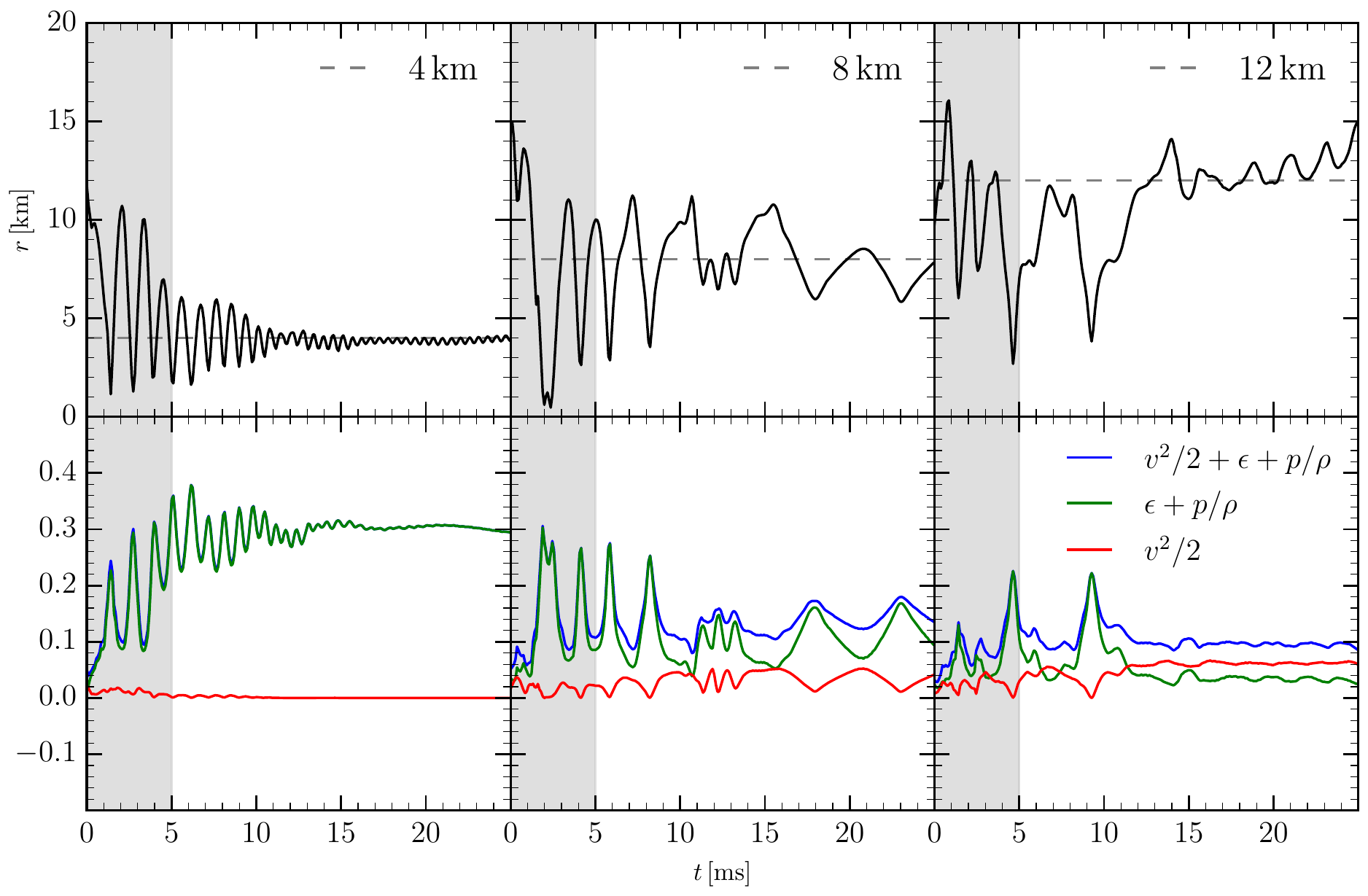}
  \caption{\textit{Top panels:} evolution of the radial positions for
    tracers in the \texttt{LS220-M132} binary that are eventually either
    in the inner regions of the HMNS (left panel), or at some distance
    from the rotation axis (central panel), or in outer regions of the
    HMNS (right panel). \textit{Bottom panels:} evolution of the
    classical Bernoulli constant \eqref{eq:bern3} relative to these
    tracers (blue solid lines), and its main contributions. The
    gray-shaded area refers to the post-merger transient when the HMNS is
    far from an equilibrium.}
\label{fig:bernoulli}
\end{center}
\end{figure*}

In its classical limit, Eq. \eqref{eq:bern0} becomes
\begin{equation}
\label{eq:bern1}
\left(1+\epsilon+\frac{p}{\rho}\right) \left( 1 + \phi +
\frac{1}{2}\vec{\boldsymbol{v}}^2 \right) =
{\rm const.}\,,
\end{equation}
where $\phi$ is the gravitational potential and where $\boldsymbol{v}$ is
the local fluid velocity. When neglecting higher-order terms, expression
\eqref{eq:bern1} further reduces to
\begin{equation}
\label{eq:bern2}
\left(
\frac{1}{2}\vec{\boldsymbol{v}}^2 + \phi + \epsilon + \frac{p}{\rho}
\right) =
{\rm const.}\,,
\end{equation}
which coincides with the classical expression for the Bernoulli constant
\cite{Rezzolla_book:2013}. Next, to translate into a classical Newtonian
language our assumption on the existence of a timelike Killing vector we
can take the gravitational potential to be independent of time and essentially
constant across the HMNS, so that Bernoulli's theorem effectively reduces
to the well-known condition that, along a fluidline,

\begin{equation}
\label{eq:bern3}
\left(
\frac{1}{2}\vec{\boldsymbol{v}}^2 + \epsilon + \frac{p}{\rho}
\right) =
{\rm const.}\,.
\end{equation}

To validate whether or not the classical Bernoulli constant
\eqref{eq:bern3} is actually a constant along a fluidline we have
calculated it for a number of tracer particles and show it for three
representative fluidlines in Fig. \ref{fig:bernoulli} for the
\texttt{LS220-M132} binary. These tracers have been selected because they
are originally in the equatorial plane and have an essentially zero
velocity in the vertical rotation, hence representing particles that are
genuinely moving in the equatorial plane. The three panels in the top row
of Fig.~\ref{fig:bernoulli} show the evolution of the radial positions
for tracers that at the end of the simulation are either in the inner
regions of the HMNS [\ie $r(t=t_{\rm fin})=4\km$, left panel], or at some
distance from the rotation axis [\ie $r(t=t_{\rm fin})=8\,{\rm km}$,
  central panel], or in outer regions of the HMNS [\ie $r(t=t_{\rm
    fin})=12\km$, right panel]. The gray-shaded area refers to the
post-merger transient when the HMNS is far from an equilibrium. Note that
these selected particles can experience large excursions from their
original positions (shown as dashed horizontal lines) due to the complex
motion around the core per Fig.~\ref{fig:tracers}, but that on average
they do not stride too far away.

\begin{figure*}
\begin{center}
\includegraphics[width=1.75\columnwidth]{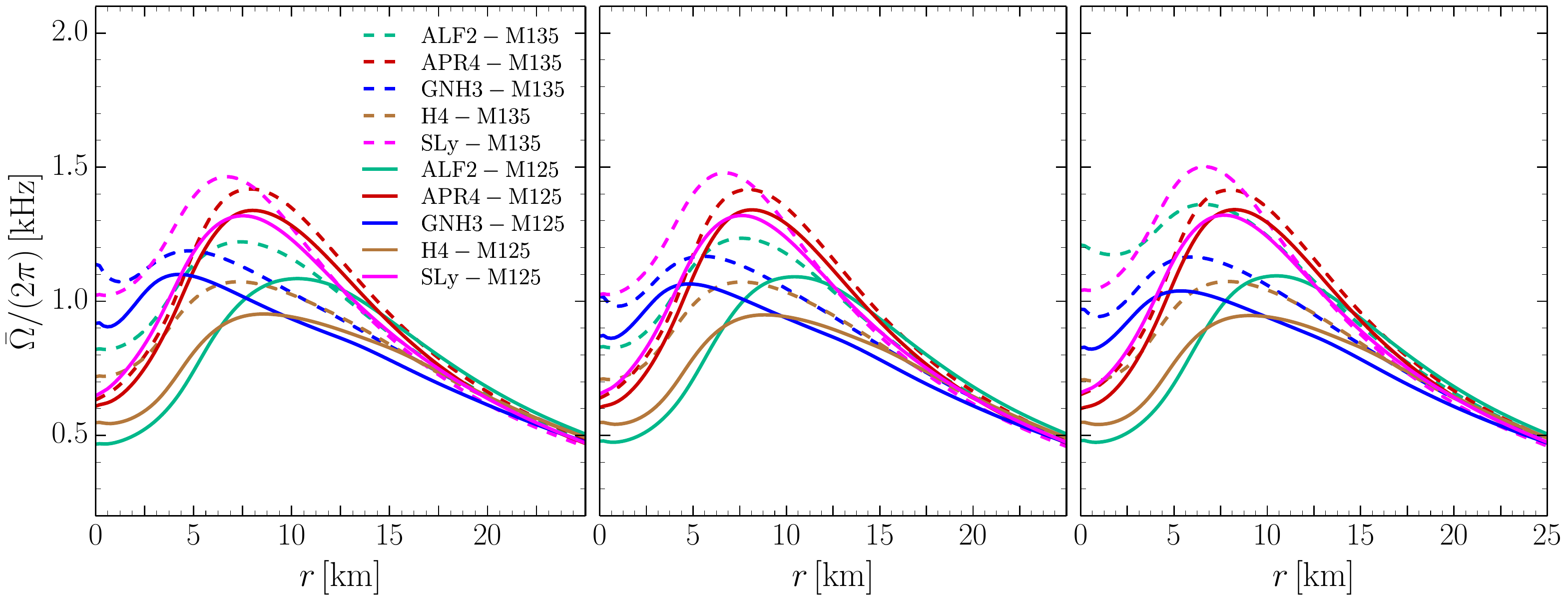}\\
\vskip 0.2cm
\includegraphics[width=1.75\columnwidth]{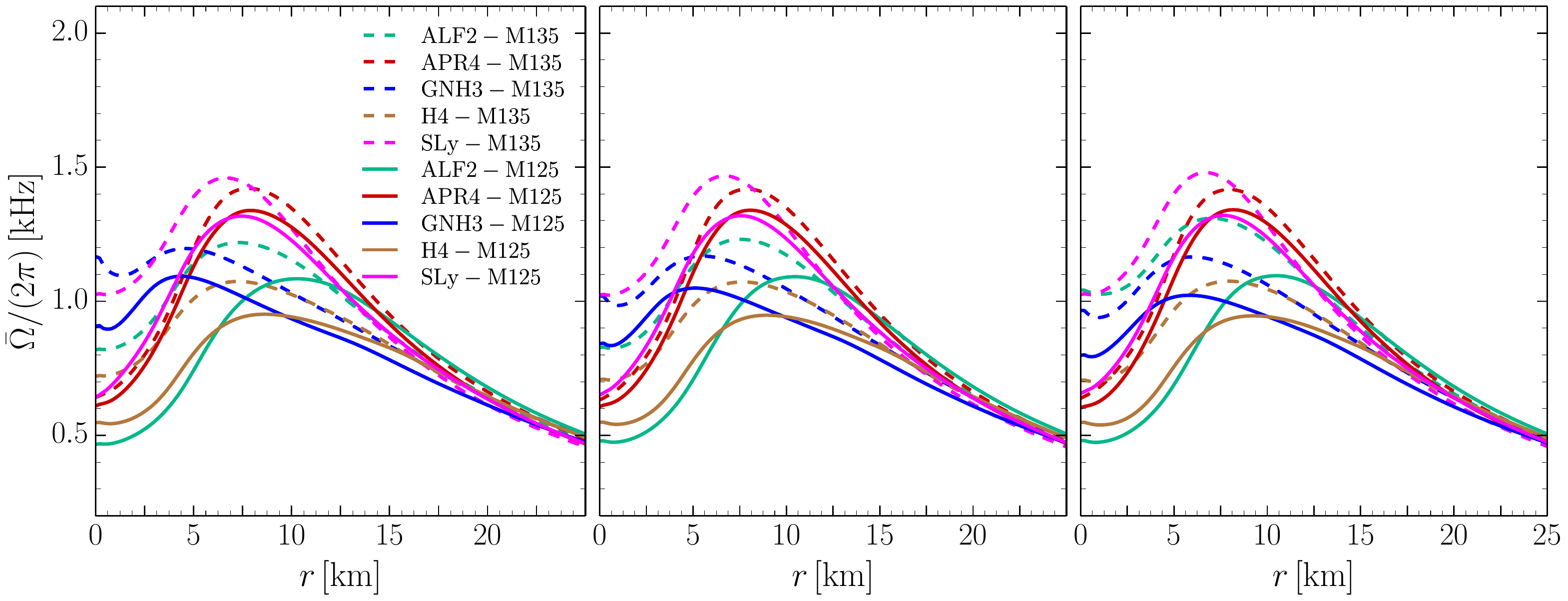}
\caption{Comparison of different averaging techniques for the time- and
  azimuthally averaged angular-velocity profiles for different EOSs. The
  beginning of the averaging window has been fixed at $6\ms$. \textit{Top
    panels:} For all EOSs and masses, the different panels refer to
  different lengths of the window, namely, $5,\,7,\,9\ms$, so that the
  from left to right the data refer to time windows $[6\,,11],\,[6\,,13]$
  and $[6\,,15]\ms$, respectively. \textit{Bottom panels:} The same as
  above but when the averaging window is not the same for the different
  EOSs and masses but is determined by the gravitational-wave frequency
  $f_2=\Omega_2/(2\pi)$. As a result, from left to right the averaging
  windows are: $[12,\,18,\,24]\times 1/f_2$,
  respectively.}\label{fig:OmegaR-AB}
\end{center}
\end{figure*}

The panels in the bottom row of Fig.~\ref{fig:bernoulli} show instead the
values of the classical Bernoulli constant \eqref{eq:bern3} relative to
the corresponding tracers in the top row (blue solid lines), but also the
two main quantities contributing to it, namely: $\epsilon + p/\rho$
(green solid lines)\footnote{Note that $\epsilon \simeq p/\rho$ at all
  times and hence they are not shown separately.} and $v^2/2$ (red solid
lines). While the values of \eqref{eq:bern3} are strictly not constant in
time (especially in transient post-merger phase indicated with the
gray-shaded areas), they also do not vary significantly around the
initial values. More importantly, it is very clear that there is a phase
opposition in the evolution of the pressure term $\epsilon + p/\rho$ and
of the kinetic term $v^2/2$, so that large values of the former
correspond to low values of the latter and viceversa. This is exactly what
one would expect in the presence of a fluid satisfying Bernoulli's
theorem, hence supporting the explanation of the phase difference in the
distribution of angular velocity, recalling that $v\sim \Omega r$, and
density being a result of the conservation of the Bernoulli quantity
$\mathcal{B} = h\,u_t$. All that has been discussed above for our three
representative tracers holds true for all others that are taken in the
neighborhood of the equatorial plane.

%%%%%%%%%%%%%%%%%%%%%%%%%%%%%%%%%%%%%%%%%%%%%%%%%%%%%%%%%%%%%%%%%
\section{Time-averaging, symmetries, and resolutions}
\label{sec:appendix_b}
%%%%%%%%%%%%%%%%%%%%%%%%%%%%%%%%%%%%%%%%%%%%%%%%%%%%%%%%%%%%%%%%%

In this Appendix we consider the impact that the time-averaging
techniques, the use of a $\pi$-symmetry and the chosen spatial resolution
have on the robustness of our results. We recall that all the simulations
reported here have used six refinement levels and a rather high spatial
resolution, namely, $\Delta h_5 = 0.15\,M_\odot \approx 221\km$ on the
finest refinement level. Furthermore, to reduce computational costs, we
have employed a reflection symmetry across the $z=0$ plane and for most
simulations a $\pi$-symmetry condition across the $x=0$ plane.

\subsection*{Impact of time-averaging techniques}

\begin{figure*}
\begin{center}
\includegraphics[width=0.99\textwidth]{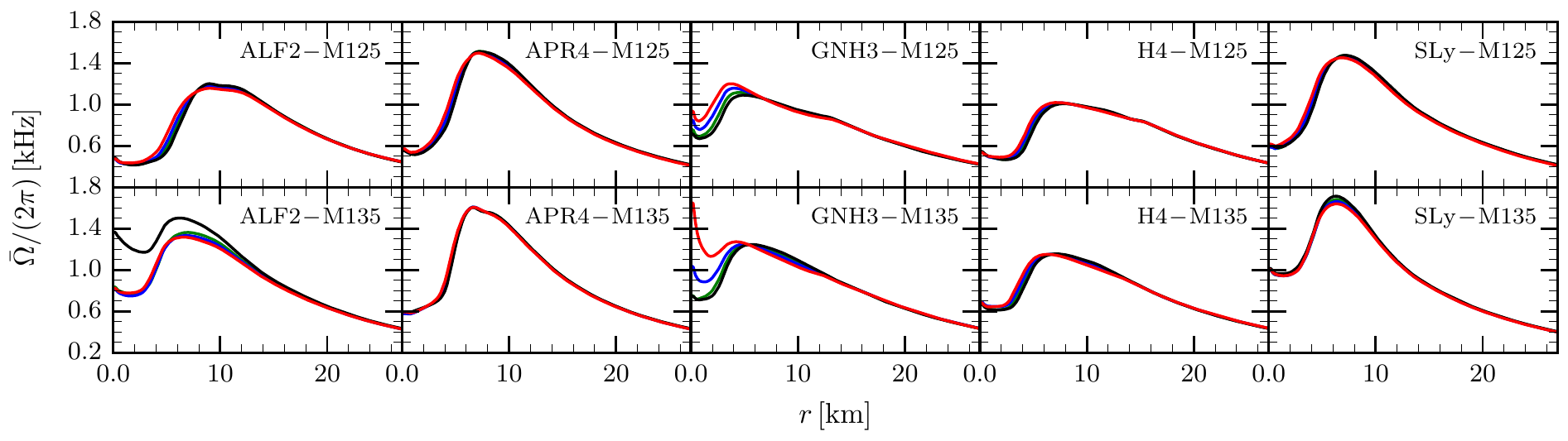}
\caption{Averaged angular-velocity profiles when the avering windows is
  set to be $7\ms$ for all EOSs and masses, but where the initial
  averaging time is varied and set to be $5$ (red line), $6$ (blue line),
  $7$ (green line), and $8\ms$ (black line), respectively. The four lines
  refer to averaging windows given by $[5, 12]$, $[6, 13]$, $[7, 14]$,
  and $[8, 15]\ms$, respectively; note that the top part of each panel
  refers to the low-mass binary, while the bottom one to the high-mass
  one.}\label{fig:omega_eos_diff_tavg}
\end{center}
\end{figure*}

The choice of the origin and length of the time-averaging window has been
guided by two principal considerations: avoiding the initial post-merger
phase and avoiding the phase briefly preceding the collapse to a black
hole. Avoiding the initial post-merger phase is important because the
HMNS is rapidly changing in its attempt to reach an equilibrium; the
matter dynamics in this phase is quite irregular, as can be seen in the
gravitational waves \cite{Rezzolla2016}, and differs significantly from
the evolution at later times when the system reaches a more equilibrium
state. On the other hand, avoiding the stage preceding the collapse to a
black hole is important because again in such a stage the dynamics is far
from equilibrium and any information on the angular velocity does not
reflect a quasi-stationary solution.

Within these constraints, there are two free parameters in performing the
time averages: the initial time of the averaging window and its width. We
recall that in the previous section we have chosen a time-averaging
centered about $t_{\rm fin}/2$ with a width of $t_{\rm fin}/3$ where
$t_{\rm fin}$ is the time to collapse to a black hole in the high-mass
cases and the end of the simulation in the low-mass cases. These values
ensure that the above mentioned considerations are realised and have no
influence on the angular-velocity profiles. Although these constraints
may appear restrictive, we consider next different times and how they
influence the averaging procedure.

To this end, we have chosen several initial and final values of the
averaging procedure and plotted the results in Fig.~\ref{fig:OmegaR-AB}.
More precisely, the top panels of Fig.~\ref{fig:OmegaR-AB} report the
angular-velocity profiles when the initial time of the averaging window
is taken to be $6 \ms$ after the merger, independently of the EOS and
mass of the binary. Furthermore, the different panels from left to right
refer to different lengths of the window, namely, $5,\,7,\,9\ms$, so that
the windows data refer to time windows $[6\,,11],\,[6\,,13]$ and
$[6\,,15]\ms$, respectively. Clearly, independent of the window length,
the qualitative features are essentially identical, exhibiting a slowly
rotating core, followed by an increase to a maximum, followed by a
decrease to a  flow {with $\Omega(r) \propto
  r^{-3/2}$}.

The bottom panels of Fig.~\ref{fig:OmegaR-AB}, on the other hand, show a
similar information in that the initial time is still fixed to $6\ms$,
but the averaging window is not the same for the different EOSs and
masses. Rather, it is determined by the gravitational-wave frequency
$f_2=\Omega_2/(2\pi)$, which is related to the maximum of the angular
velocity $\Omega_{\rm max}$ (\cf Fig. \ref{fig:OmegaO2_3p}). In this
way, each binary will have an average window which is set to be a
multiple of the spinning frequency of the HMNS. In practice we have set
the averaging window to be $\Delta t = [12,\,18,\,24]\times 1/f_2$ in the
panels from left to right, respectively. Also in this case, the
qualitative behaviour of the various angular-velocity profiles is the
same and the differences are of a few percent at most.

As a final variant of the possible way of performing the time averages,
we report in Fig.~\ref{fig:omega_eos_diff_tavg} the angular-velocity
profiles when the averaging windows is set to be $7\ms$ for all EOSs and
masses, but where the initial averaging time is varied and set to be
$5,\,6,\,7$ and $8\ms$, respectively. As a result, the four lines
reported in each panel refer to averaging windows given by $[5, 12]$,
$[6, 13]$, $[7, 14]$, and $[8, 15]\ms$, respectively; note that the top
part of each panel refers to the low-mass binary, while the bottom one to
the high-mass binary. Also when considering this different technique it
emerges rather clearly that the averaging procedure has little influence
on the angular-velocity distribution. However, two exceptions are also
equally clear and for obvious reasons. The first one is offered by the
binary \texttt{ALF2-M135} case, whose HMNS collapses at approximately
$15\ms$ (\cf Fig.~\ref{fig:GW-ALF2-M135}) and whose ``late-time''
averaging window is obviously spoiled by the large increase in $\Omega$
occurring before the collapse. The second exception is given instead by
the binary \texttt{GNH3-M135}, which has instead a long-lasting transient
post-merger phase, with the two stellar cores still clearly visible. Also
in this, the ``early-time'' averaging is not representative of a
quasi-stationary stage. Excluding these two obvious pathological
averaging windows, the maximum angular-velocity changes by $5\%$ at most
for all masses and EOSs.

\subsection*{Impact of $\pi$-symmetry} 

\begin{figure*}
\begin{center}
\includegraphics[width=\columnwidth]{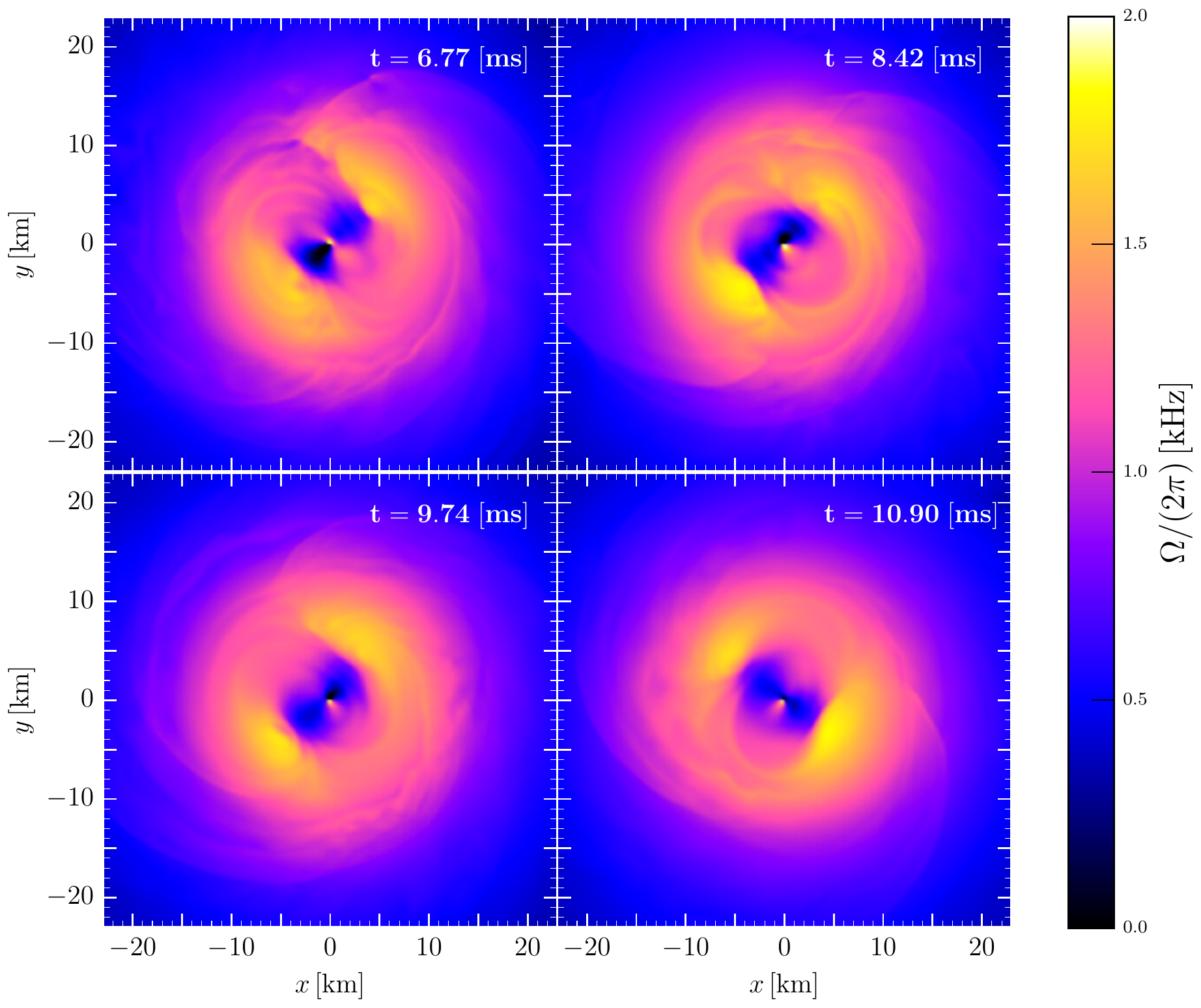}
\hskip 0.5cm
\includegraphics[width=\columnwidth]{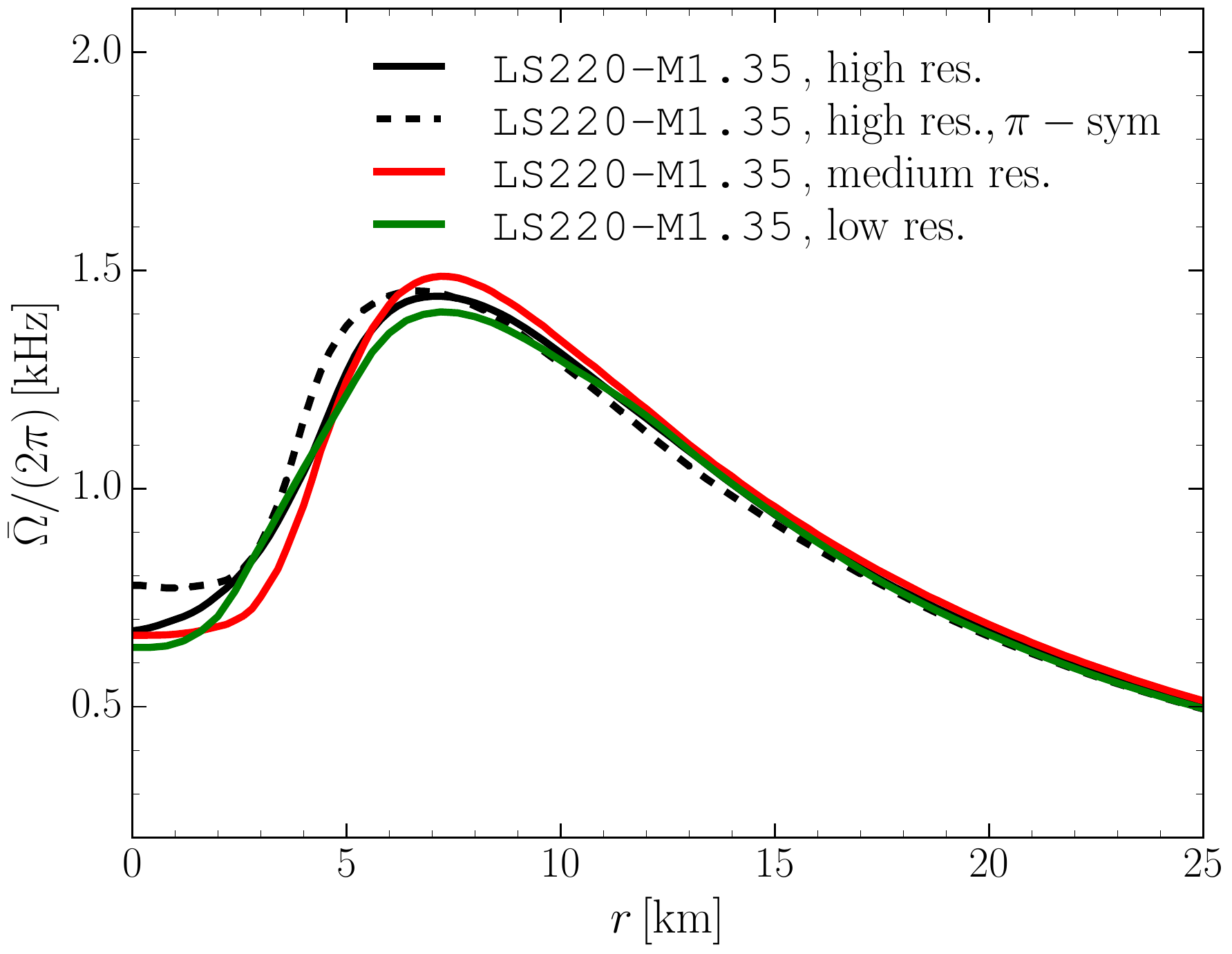}
\caption{\textit{Left panel:} angular velocity distribution on the
  equatorial plane at four representative times for a binary with the
  LS220 EOS evolved without $\pi$-symmetry; note the appearance of an
  $m=1$ deformation in addition to the larger $m=2$ deformation.
  \textit{Right panel:} corresponding azimuthal and time-averaged profile
  for the same binary with $\pi$-symmetry (black dashed line) and without
  (black solid line) at a resolution of $\Delta x =0.15$. Additionally,
  low resolution runs of $\Delta x=0.20$ (red solid line) and $\Delta x
  =0.25$ (green solid line) are shown.  All resolutions exhibit the same
  behaviour already discussed above.}\label{fig:meq1}
\end{center}
\end{figure*}

We next consider the impact of having imposed a $\pi$-symmetry in our
simulations. While this is a perfectly reasonable option in view of the
considerable savings in computational costs, it also blinds us to the
development of an $m=1$ instability that has been reported by a number of
groups \cite{East2016, Radice2016a, Lehner2016a}. While the
gravitational-wave signal associated with the instability is always
smaller than the dominant one coming from the $m=2$ deformations in the
HMNS, so that its observation by current generation detectors is unlikely
and will require third-generation detectors \cite{Radice2016a}, it is
useful to verify whether the presence of the one-arm instability would
leave an imprint on the angular-velocity profiles despite the azimuthal
average.

To this scope we have considered the evolution of an equal-mass binary
with the LS220 EOS and a gravitational mass of $2\times 1.350\,M_{\odot}$
(\cf binary \texttt{LS220-M135} in Table \ref{tab:models}), evolved with
and without $\pi$-symmetry to investigate the influence of the
instability on the rotation profiles. The corresponding angular-velocity
distribution on the equatorial plane for the simulation without the
$\pi$-symmetry is shown in the left panel of Fig.~\ref{fig:meq1} at four
representative times after the merger and when the HMNS has reached a
quasi-stationary state. Comparing such a panel with the bottom rows of
Figs.~\ref{fig:ALF2-M135-rho} and \ref{fig:ALF2-M125-rho}, where the
$\pi$-symmetry is imposed, highlights the presence of a small $m=1$
deformation. The right panel of Fig.~\ref{fig:meq1}, on the other hand,
reports the corresponding azimuthal and time-averaged profile for two
simulations. The black dashed line refers to the $\pi$-symmetric run,
while the black solid line to the run without $\pi$-symmetry; in both
cases the average is done between $t=5\ms$ and $t=12\ms$.

Clearly, no sign of the $m=1$ deformation is present, as one would expect
from an averaging process; rather, the angular velocity shows similar
quantitative behaviour with and without the use of $\pi$-symmetry. The
greatest difference is in the very centre of the HMNS where the angular
velocity is higher with $\pi$-symmetry than without. The maximum angular
velocity is $1$\% larger with $\pi$-symmetry and the location of the
maximum is slightly shifted to larger radii. Both runs exhibit
 {quasi-circular} orbits at larger radii. We
conclude that in the very interior of the HMNS, the use of $\pi$-symmetry
plays a small role, but also that outside a core region of $\simeq 5\km$
the influence is minimal and does not affect our conclusions.

\subsection*{Impact of grid resolution} 

Since the stability properties of the HMNS phase depend on the resolution
(see \cite{Baiotti:2009gk} where this was first investigated
systematically), the determination of its lifetime against gravitational
collapse requires a systematic and very careful resolution study. At the
same time, because of the development of large shocks, the convergence
order after the merger is inevitably very low (\ie of order unity or
less), so that a strict convergence study of this stage is of little use
and certainly beyond the scope of this paper. In fact, different
resolutions would mostly produce phase differences in the dynamics of the
fluid and spacetime variables, hence with only a small impact on our
results that are expressed in terms of time and azimuthal averages.

Notwithstanding these considerations, it is reasonable to ask how
significant are the changes in the angular-velocity profiles when the
simulations are performed at different resolutions. Such a resolution
study would then provide confidence on the robustness of the results
presented here. We also recall that in Ref. \cite{Takami2015} we have
considered the influence of the resolution on the dynamics of the HMNS by
using three different resolutions for a binary described by an
ideal-fluid EOS. As remarked in \cite{Takami2015}, the rather high
resolution employed here on the finest refinement level, \ie $\Delta h_5
= 0.15\,M_{\odot}$, provides a description of the HMNS which is very
close to that obtained with an even higher resolution of $\Delta h_5 =
0.125\,M_{\odot}$.

The results of our resolution study are summarized in the right panel of
Fig.~\ref{fig:meq1} for the \texttt{LS220-M135} binary. More
specifically, in the right panel of Fig.~\ref{fig:meq1} we plot the
averaged angular velocity for the binary \texttt{LS220-M135} for three
different resolutions on the finest refinement level, \ie $\Delta
h_5=0.15,0.20,0.25\,\Msun \approx 221, 295, 369\,{\rm m}$, and which we
dub as ``high'', ``medium'' and ``low resolution'', respectively. All
simulations do not use a $\pi$-symmetry, except where noted. We also
remark that although a resolution of $\Delta x=0.25\,\Msun$ may appear
coarse, it is routinely used in numerical-relativity simulations of
binary neutron stars (see, \eg \cite{DePietri2016, Maione2016, Feo2017})
and has been shown to be high enough to provide physically robust results
(see \cite{DePietri2016} for an extensive discussion).

Clearly, at all resolutions the profile of the angular-velocity is
similar, namely, showing a slowly rotating core, rising to a maximum
around $8\km$ before decreasing to a {$r^{-3/2}$} profile. The largest
differences between resolution are in the centre of the HMNS, where the
rest-mass densities are the highest and the metric functions show the
largest gradients. Despite this, all resolutions reach a maximum
angular-velocity at around $8\km$ with a variation with resolution that
is at most $5\%$. This small variance demonstrates the robustness of the
maximum angular-velocity and illustrates that quasi-universal relations
proposed in Sec. \ref{sec:qub} are a robust feature of the HMNS.

%%%%%%%%%%%%%%%%%%%%%%%%%%%%%%%%%%%%%%%%%%%%%%%%%%%%%%%%%%%%%%%%%%
%%%   REFERENCES
%%%%%%%%%%%%%%%%%%%%%%%%%%%%%%%%%%%%%%%%%%%%%%%%%%%%%%%%%%%%%%%%%%

\bibliographystyle{apsrev4-1}

\end{document}